%
%
%
%
\magnification\magstep 1
\parskip 4pt plus 1pt minus 0.5pt
\def\d{{\rm d}}
\def\e{{\bf e}}
\def\J{{\cal I}}
\def\n{{\bf n}}

\def\p{{\bf p}}
\def\q{{\bf q}}
\def\r{{\bf r}}
\def\R{R_{\rm eff}}
\def\s{\sigma}
\def\x{{\bf x}}
\def\y{{\bf y}}
\def\z{\zeta}
\def\cot{\mathop{\rm cot}}
\def\crr{\cr\noalign{\hrule}}
\def\dif{^{\rm diff}}
\def\dint{\int\!\!\!\!\!\int}
\def\euler{\gamma_{\scriptscriptstyle E}}
\def\frac#1#2{{#1\over #2}}
\def\frad#1#2{\displaystyle{#1\over #2}}
\def\full{_{\rm full}}
\def\init{\tabskip 0pt}
\def\re{\mathop{\rm Re}}
\def\sp{{\ \ }}
\def\stot{\sigma_{\rm tot}}
{\parskip 0pt\parindent 0pt
\centerline{\bf LIGHT SCATTERING FROM MESOSCOPIC OBJECTS}
\medskip
\centerline{\bf IN DIFFUSIVE MEDIA}
\bigskip\null\medskip
\centerline{by J.M. Luck$^{(1,\rm a)}$ and Th.M. Nieuwenhuizen$^{(2,\rm b)}$}
\bigskip
(1) C.E.A. Saclay, Service de Physique Th\'eorique,
91191 Gif-sur-Yvette cedex, France.
\medskip
(2) Van der Waals-Zeeman Laboratorium,
Valckenierstraat 65, 1018 XE Amsterdam, The Netherlands.
\vfill
{\bf Abstract.}
No direct imaging is possible in turbid media,
where light propagates diffusively over length scales larger than
the mean free path $\ell$.
The diffuse intensity is, however, sensitive to the presence
of any kind of object embedded in the medium, e.g. obstacles or defects.
The long-ranged effects of isolated objects in an otherwise homogeneous,
non-absorbing medium can be described by a stationary diffusion equation.
In analogy with electrostatics, the influence of a single embedded object
on the intensity field is parametrized in terms of a multipole expansion.
An absorbing object is chiefly characterized by a negative charge,
while the leading effect of a non-absorbing object
is due to its dipole moment.
The associated intrinsic characteristics of the object
are its capacitance $Q$ or its effective radius $\R$,
and its polarizability $P$.
These quantities can be evaluated within the diffusion approximation
for large enough objects.
The situation of mesoscopic objects,
with a size comparable to the mean free path,
requires a more careful treatment,
for which the appropriate framework is provided by radiative transfer theory.
This formalism is worked out in detail,
in the case of spherical and cylindrical objects of radius $R$,
of the following kinds:
(i) totally absorbing (black), (ii) transparent, (iii) totally reflecting.
The capacitance, effective radius, and polarizability
of these objects differ from the predictions of the diffusion approximation
by a size factor, which only depends on the ratio $R/\ell$.
The analytic form of the size factors is derived for small and large objects,
while accurate numerical results are obtained
for objects of intermediate size $(R\sim\ell)$.
For cases~(i) and (ii)
the size factor is smaller than one and monotonically increasing with $R/\ell$,
while for case~(iii) it is larger than one and decreasing with $R/\ell$.
\vfill
P.A.C.S.: 42.30.-d, 42.25.Bs, 42.68.Ay
\smallskip
(a) e-mail: luck@spht.saclay.cea.fr

(b) e-mail: nieuwenh@phys.uva.nl
}
\eject
\noindent{\bf 1 INTRODUCTION}
\smallskip

Light undergoes multiple scattering when propagating through
an inhomogeneous medium over distances much larger than one mean free path
(see ref.~[1] for a review).
Objects embedded in such a turbid medium are invisible from the outside:
no ballistic imaging is possible.
Obstacles or defects can, however, be detected by a careful analysis
of the perturbations of the diffuse intensity field.
Imaging through turbid media has been the subject of many investigations.
It has a wide range of applications,
especially in the field of medical imaging.
Just to mention one example,
at this moment the Philips Laboratories in Eindhoven (the Netherlands)
are testing prototype devices to detect breast cancer
by diffuse light propagation.
This method is non-invasive:
as opposed to conventional X-ray tomography,
it does not cause radiation damage,
even if applied to a large population over a long period of time.

In ref.~[2] it has been shown theoretically, and demonstrated experimentally,
that the diffuse image of an object embedded in a homogeneous turbid medium,
whether absorbing or not,
can be quantitatively described by the diffusion approximation.
This framework, to be recalled in section~2,
is advantageously rephrased in the language of electrostatics.
The long-distance effect of an embedded object on the diffuse
intensity field $I(\x)$ is characterized by a multipole expansion.
If the object absorbs radiation, the leading term is a negative charge $q$,
while a non-absorbing object is chiefly described by a dipole moment $\p$.
Like in electrostatics, higher-order multipoles are most often negligible.
This framework leads
to a quantitative prediction of the diffuse image of the object,
i.e., the profile of diffuse intensity transmitted through a thick slab
near the embedded object.
In ref.~[2] the charge and the dipole moment of a sphere and of a cylinder
have been calculated within the diffusion approximation,
namely by solving the diffusion equation in the presence of a
spherical or cylindrical object.
This approach is justified a priori for {\it macroscopic} objects,
with a size much larger than the mean free path $\ell$.
Further results on imaging of objects within the framework
of the diffusion approximation can be found in refs.~[3--6].

The diffusion approximation breaks down in the case of
{\it mesoscopic} embedded objects,
with a size comparable to the mean free path.
Recently, Lancaster and Nieuwenhuizen have investigated the cases
of point scatterers and of small mesoscopic scatterers~[7].
In the situation of point scatterers, the analysis is rather simple,
although short-distance singularities have to be regularized,
introducing thus new effective parameters.
A partial-wave expansion was introduced
in the more difficult case of extended objects.

The goal of the present work is to propose a more general approach
to the quantitative calculation
of the charge and dipole of a mesoscopic object.
We shall put the emphasis on the associated intrinsic characteristics:
capacitance $Q$, effective radius $\R$, polarizability $P$.
The natural framework to evaluate these parameters
is radiative transfer theory (RTT)~[8--11],
at least in the semi-classical regime,
corresponding to most experimental situations, defined by $\ell\gg\lambda$,
with $\lambda$ being the wavelength of radiation.
RTT provides a quantitative description of the intensity over
mesoscopic length scales, down to the mean free path $\ell$,
whereas the diffusion approximation only describes the variations
of the diffuse intensity $I(\x)$ over distances much larger than $\ell$.
This advantage of RTT with respect to the diffusion approximation,
especially when dealing with skin-layer phenomena near boundaries,
has been emphasized in several recent works~[12--14].
For definiteness and simplicity,
we consider multiple isotropic scattering of scalar waves
in a non-absorbing medium.
We focus our attention on spherical or
cylindrical embedded objects, with radius $R$,
and restrict the analysis to the following three cases:

\noindent(i) totally absorbing (black) object,
{\parskip 0pt

\noindent(ii) transparent object, with the same optical index as the medium,

\noindent(iii) totally reflecting object.
}

\noindent In section~2 we recall the diffusion approach of ref.~[2].
In particular we express the charge and the dipole of an embedded object
in terms of intrinsic characteristics,
its capacitance $Q$ or its effective radius $\R$, and its polarizability $P$,
and we recall the predictions of the diffusion approximation in the cases
listed above.
In section~3 we employ RTT in order to derive the capacitance $Q$
or the polarizability $P$ of a sphere in the same three cases.
Analytical predictions are obtained for small spheres $(R\ll\ell)$
and large spheres $(R\gg\ell)$,
while the appropriate RTT equation is solved numerically
in the general case $(R\sim\ell)$.
The same analysis is then performed in section~4
for the effective radius $\R$ or the polarizability $P$ of cylindrical objects.
Section~5 contains a brief discussion.

\medskip
\noindent{\bf 2 GENERAL FRAMEWORK}
\smallskip
\noindent{\bf 2.1 Multipole expansion}
\smallskip

Our goal is to investigate the diffuse image of a small object
embedded in a homogeneous, non-absorbing multiple-scattering medium.
The size $L$ of the sample is assumed to be very large
with respect to the mean free path $\ell$,
so that the diffusion approximation can be used to describe
the intensity $I(\x)$,
everywhere except in a {\it skin layer} around the object under study,
with a thickness of order $\ell$.
The intensity $I(\x)$ is therefore the solution
of the stationary diffusion equation
$$
\nabla^2 I(\x)=0,
\eqno(2.1)
$$
with appropriate boundary conditions,
taking account of sources, of the geometry of the sample,
and of the presence of the embedded object.
Let $I_0(\x)$ be the solution of eq.~(2.1) in the absence of the object.
The total intensity reads
$$
I(\x)=I_0(\x)+\delta I(\x).
\eqno(2.2)
$$
The main goal of the present work is to provide a quantitative characterization
of the disturbance $\delta I(\x)$ of the diffuse intensity,
representing the diffuse image of the embedded object.

\medskip
\noindent$\bullet$ {\it The case of a sphere}
\smallskip

Consider first the case where the embedded object is a small sphere,
with radius $R$, located at the position $\x_0$ in the medium.
It is expected on general grounds that the disturbance $\delta I(\x)$
can be characterized by a multipole expansion,
in analogy with electrostatics~[2,~4--7].
It turns out that only the first two of them,
namely the {\it charge} $q$ and the {\it dipole moment} $\p$,
play a role, while higher-order multipoles are most often negligible.
Assume for a while that the medium is infinite.
Setting $\r=\x-\x_0$ and $r=\vert\r\vert$,
the disturbance of the intensity far enough from the object assumes the form
$$
\delta I(\x)=(q+\p\cdot\nabla)\frac{1}{r}=\frac{q}{r}-\frac{\p\cdot\r}{r^3}.
\eqno(2.3)
$$
This expression, which is a solution of the diffusion equation~(2.1),
holds as soon as the distance to the object
is large with respect to the mean free path $\ell$.

The linearity of the problem and the isotropy of the spherical defect
imply that the charge $q$ is proportional to the local intensity
$I_0(\x_0)$ in the absence of the object,
while the dipole moment $\p$ is proportional to its gradient.
We therefore set
$$
q=-QI_0(\x_0),\quad\p=-P\nabla I_0(\x_0),
\eqno(2.4)
$$
and we propose to call $Q$ the {\it capacitance} of the spherical object,
and $P$ its {\it polarizability}.
These two parameters are intrinsic characteristics of the object.
Their dependence on its radius and on its optical properties
will be investigated in detail in the following.

The capacitance $Q$ is non-zero (and in fact positive)
only if the sphere is absorbing.
Indeed, eq.~(2.3) implies that the total flux of radiation
absorbed by the object reads
$$
\delta F=-4\pi q=4\pi QI(\x_0).
\eqno(2.5)
$$
For a non-absorbing object,
the leading contribution is that of its dipole $\p$.
The corresponding polarizability $P$ can be either positive or negative,
depending on the nature of the sphere.
We finally observe that $Q$ has the dimension of a length,
while $P$ has the dimension of a volume.
As a consequence, for a sphere of radius $R$,
these parameters scale as $Q\sim R$ and $P\sim R^3$, roughly speaking.
More accurate predictions will be derived throughout the rest of this paper.

\medskip
\noindent$\bullet$ {\it The case of a cylinder}
\smallskip

The present formalism can be extended to other objects.
Let us consider the case of a rod, modeled as an infinitely long cylinder,
with radius $R$, whose axis is parallel to the $z$-axis.
Assume that the sample is infinitely long in this direction.
Then both $I_0(\x)$ and the disturbance $\delta I(\x)$
only depend on the two-dimensional perpendicular component $\x_\perp=(x,y)$.
Eq.~(2.3) is replaced by
$$
\delta I(\x_\perp)=(q+\p\cdot\nabla)\ln\frac{r_0}{r}
=q\ln\frac{r_0}{r}-\frac{\p\cdot\r_\perp}{r^2},
\eqno(2.6)
$$
where $\r_\perp=\x_\perp-(\x_0)_\perp$, so that $r=\vert\r_\perp\vert$
is the distance to the axis of the cylinder.
The length scale $r_0$, requested by dimensional analysis,
is determined by conditions at the boundaries of the sample,
so that it is in general proportional to the sample size.
This sensitivity of $\delta I(\x)$ to global properties
of the sample arises because the logarithmic potential of a point charge
in two-dimensional electrostatics is divergent at long distances.
As a consequence, the capacitance $Q$ is not intrinsic to the cylinder.
An intrinsic quantity is its {\it effective radius} $\R$,
defined by the condition that the cylindrically symmetric part
of the total intensity, including the charge term in the right-hand side
of eq.~(2.6), vanishes at a distance $r=\R$ from the axis of the cylinder.
We thus have $I_0(\x_0)+q\ln(r_0/\R)=0$, hence
$$
Q=\frac{1}{\ln\frad{r_0}{\R}}.
\eqno(2.7)
$$
The general remarks of the previous paragraph also apply to a cylinder.
We observe that $\R$ and $r_0$ have the dimension of a length,
while the capacitance $Q$ is dimensionless,
and the polarizability $P$ has the dimension of an area.
As a consequence, the orders of magnitude $\R\sim R$ and $P\sim R^2$
can be expected for a cylinder of radius $R$.
More accurate predictions will be derived throughout the rest of this paper.

\medskip
\noindent{\bf 2.2 Transmission in a thick-slab geometry}
\smallskip

In order to illustrate the definitions of the capacitance $Q$,
of the effective radius $\R$, and of the polarizability $P$
given in section~2.1,
we recall the derivation of the diffuse image of a small object,
i.e., the profile of transmitted intensity, through a thick-slab sample.
This geometry is often encountered experimentally~[2].

We consider an optically thick slab $(0<x<L)$, with $L\gg\ell$,
assumed to be infinite in the two transverse directions.
The left side of the sample $(x=0)$ is subjected to an incident plane wave,
so that the solution of eq.~(2.1) when the object is absent reads
$$
I_0(\x)=I_0\left(1-\frac{x}{L}\right).
\eqno(2.8)
$$

The intensity $T(y,z)$ transmitted through the sample,
and emitted on the right side $(x=L)$ at the point $(y,z)$,
is proportional to the normal derivative of the diffuse intensity
at that point:
$$
T(y,z)=-K\ell\left.\frac{\partial I(x,y,z)}{\partial x}\right\vert_{x=L}.
\eqno(2.9)
$$
When there is no embedded object, eq.~(2.8) yields a uniform transmission
$$
T_0=\frac{KI_0\ell}{L}.
\eqno(2.10)
$$
The prefactor $K$ bears the dependence of the transmitted intensity
on the angle and polarization of the incident and emitted radiation.
If these characteristics are not resolved,
namely if the ``all-in, all-out'' intensity is measured,
the prefactor $K$ assumes the value $K=4\pi/(3\lambda^2)$,
with $\lambda$ being the wavelength of radiation.
This universal result is analogous
to the Boltzmann formula for the conductivity~[12,~1].
The dependence of $K$ on incidence angles and polarizations is non-trivial,
as it involves microscopic characteristics of the medium,
such as the anisotropy of the scatterers.
These features cannot be accurately described
within the diffusion approximation.
Indeed, they originate in the presence of skin layers
near the boundaries of the sample, with a thickness comparable to $\ell$,
where the free radiation is converted into diffuse intensity, and vice versa.
The appropriate tool to investigate skin-layer phenomena
is the Schwarzschild-Milne equation of RTT~[12--14].
Ref.~[1] contains an overview of these recent works.

\medskip
\noindent$\bullet$ {\it The case of a sphere}
\smallskip

Consider first the case where the embedded object is a sphere,
with capacitance $Q$ and polarizability $P$, located at $x=x_0$, $y=z=0$.
Its charge and dipole moment read
$$
q=-QI_0\left(1-\frac{x_0}{L}\right),\quad\p=p\e_x,\quad p=\frac{PI_0}{L},
\eqno(2.11)
$$
with $\e_x$ being the unit vector of the positive $x$-axis.

Along the lines of ref.~[2],
the total intensity can be calculated by summing the infinite-space
expression~(2.3) over a periodic double array of images.
Setting $\rho=(y^2+z^2)^{1/2}$, we obtain
$$
\eqalign{
I(\x)=I_0\left(1-\frac{x}{L}\right)
&+q\!\sum_{n=-\infty}^{+\infty}\!
\left(\frac{1}{\big[(x-x_0+2nL)^2+\rho^2\big]^{1/2}}
-\frac{1}{\big[(x+x_0+2nL)^2+\rho^2\big]^{1/2}}\right)\cr
&-p\!\sum_{n=-\infty}^{+\infty}\!
\left(\frac{x-x_0+2nL}{\big[(x-x_0+2nL)^2+\rho^2\big]^{3/2}}
+\frac{x+x_0+2nL}{\big[(x+x_0+2nL)^2+\rho^2\big]^{3/2}}\right).\cr
}
\eqno(2.12)
$$

The transmitted intensity $T(y,z)=T(\rho)$ then reads
$$
\eqalign{
T(\rho)=T_0\Bigg(
1&-2Q(L-x_0)\sum_{n=-\infty}^{+\infty}
\frac{L-x_0+2nL}{\big[(L-x_0+2nL)^2+\rho^2\big]^{3/2}}\cr
&-2P\sum_{n=-\infty}^{+\infty}
\frac{2(L-x_0+2nL)^2-\rho^2}{\big[(L-x_0+2nL)^2+\rho^2\big]^{5/2}}\Bigg).\cr
}
\eqno(2.13)
$$

The diffuse image of the embedded sphere thus extends over a region
of transversal size $\rho\sim L-x_0$.
If the sphere is absorbing, the leading contribution of its charge is negative,
with a relative magnitude proportional to $Q/L$, i.e., to $R/L$.
If the sphere is not absorbing,
the leading contribution of its dipole is proportional to $P/L^3$,
i.e., to $R^3/L^3$, in magnitude.
It is negative for $P>0$ (e.g. for a reflecting sphere),
and positive for $P<0$ (e.g. for a transparent sphere).

\medskip
\noindent$\bullet$ {\it The case of a cylinder}
\smallskip

Consider the case where the embedded object is a cylinder,
with given effective radius $\R$ and polarizability $P$,
whose axis is parallel to the $z$-axis, and contains the point $x=x_0$, $y=0$.
The total intensity can be calculated by applying the method of images
to eq.~(2.6).
We thus obtain
$$
\eqalign{
I(\x)=I_0\left(1-\frac{x}{L}\right)
&-\frac{q}{2}\sum_{n=-\infty}^{+\infty}
\ln\frac{(x-x_0+2nL)^2+y^2}{(x+x_0+2nL)^2+y^2}\cr
&-p\sum_{n=-\infty}^{+\infty}
\left(\frac{x-x_0+2nL}{(x-x_0+2nL)^2+y^2}
+\frac{x+x_0+2nL}{(x+x_0+2nL)^2+y^2}\right).\cr
}
\eqno(2.14)
$$
This result has the following alternative closed-form expression
$$
\eqalign{
I(\x)=I_0\left(1-\frac{x}{L}\right)
&-q\re
\left(\ln\sin\frac{\pi(\z-x_0)}{2L}-\ln\sin\frac{\pi(\z+x_0)}{2L}\right)\cr
&-p\frac{\pi}{2L}\re
\left(\cot\frac{\pi(\z-x_0)}{2L}+\cot\frac{\pi(\z+x_0)}{2L}\right),\cr
}
\eqno(2.15)
$$
in terms of the complex variable $\z=x+iy$.

The charge $q$ can be calculated by expressing that
the first line of the right-hand side of eq.~(2.15),
including the free term and the charge term,
vanishes at the distance $\R\ll L$ from the axis of the cylinder.
We thus obtain
$$
q=-QI_0\left(1-\frac{x_0}{L}\right),
\quad\hbox{with}\quad Q=\frac{1}{\ln\frad{r_0}{\R}}
\quad\hbox{and}\quad r_0=\frac{2L}{\pi}\sin\frac{\pi x_0}{L},
\eqno(2.16)
$$
in agreement with the general results~(2.4), (2.7).
This example shows that the length scale $r_0$ is proportional
to the sample thickness $L$,
and that it depends on the position of the embedded object.

Finally, the transmitted intensity $T(y,z)=T(y)$ reads
$$
T(y)=T_0\left(1-Q\pi\left(1-\frac{x_0}{L}\right)
\frac{\sin\frad{\pi x_0}{L}}{\cos\frad{\pi x_0}{L}+\cosh\frad{\pi y}{L}}
-P\frac{\pi^2}{L^2}\frac{1+\cos\frad{\pi x_0}{L}\cosh\frad{\pi y}{L}}
{\left(\cos\frad{\pi x_0}{L}+\cosh\frad{\pi y}{L}\right)^2}\right).
\eqno(2.17)
$$

The diffuse image of the cylinder again extends over a region
of transversal size $y\sim L-x_0$.
If the cylinder is absorbing,
the leading contribution of its charge is negative.
Its relative magnitude, proportional to $Q$,
depends logarithmically on the sample thickness $L$ via $r_0$,
according to eq.~(2.16).
If the cylinder is not absorbing,
the leading contribution of its dipole has a magnitude of order $P/L^2$,
i.e., $R^2/L^2$.

The characteristic profiles of transmitted radiation recalled above
have been observed experimentally by den Outer et al.~[2],
in the case of pencils (black cylinders)
and glass fibers (transparent cylinders with a non-trivial index mismatch).

\vfill\eject
\noindent{\bf 2.3 Diffusion approximation}
\smallskip

When the radius $R$ of the embedded object is much larger than $\ell$,
its characteristics $Q$ or $\R$ and $P$
can be calculated within the diffusion approximation,
along the lines of ref.~[2].
Indeed, the diffusion approximation only breaks down in a skin layer
of thickness $\ell\ll R$ around the surface of the object.
We summarize, for completeness, the main predictions of this approach,
in the three cases listed in the Introduction.

\medskip
\noindent{\sl 2.3.1 The capacitance of a totally absorbing (black) object}
\smallskip
\noindent$\bullet$ {\it The case of a sphere}
\smallskip

Consider first a totally absorbing sphere
of radius $R\gg\ell$, located at the origin in an infinite medium.
Within the diffusion approximation,
the intensity $I(\x)=I(r)$ is a solution of eq.~(2.1),
more precisely a spherically symmetric harmonic function,
with the absorbing boundary condition $I(R)=0$ at the surface of the sphere.
A basis of harmonic functions is given by $\{1, 1/r\}$,
so that the required solution is proportional to
$$
I(r)=1-\frac{R}{r}.
\eqno(2.18)
$$
By identifying this expression with eqs.~(2.3), (2.4), we get the simple result
$$
Q\dif=R
\eqno(2.19)
$$
for the capacitance of an absorbing sphere, in the diffusion approximation.
We recall that this result is expected to hold in the regime $R\gg\ell$.
The dependence of $Q$ on the ratio $R/\ell$ will be investigated in
section~3.2.

\medskip
\noindent$\bullet$ {\it The case of a cylinder}
\smallskip

Consider a totally absorbing cylinder of radius $R\gg\ell$,
whose axis is parallel to the $z$-axis and passes through the origin.
A basis of cylindrically symmetric harmonic functions is given
by $\{1,\ln r\}$, so that the required solution is proportional to
$$
I(r)=\ln\frac{R}{r}.
\eqno(2.20)
$$
We thus have
$$
\R\dif=R.
\eqno(2.21)
$$
The effective radius of an absorbing cylinder
coincides with its geometrical radius, within the diffusion approximation.
Again, this result is expected to hold for $R\gg\ell$ only,
while the dependence of $\R$ on $R/\ell$ will be investigated in section~4.2.

\vfill\eject
\noindent{\sl 2.3.2 The polarizability of a non-absorbing object}
\smallskip
\noindent$\bullet$ {\it The case of a sphere}
\smallskip

Consider a non-absorbing sphere of radius $R\gg\ell$,
located at the origin in an infinite medium.
Within the diffusion approximation,
the only property of a material which is naturally involved
is the diffusion constant of radiation in its bulk.
So, assume that the diffusion constant reads
$D_1$ in the medium and $D_2$ in the sphere.

Let the intensity be $I_1(\x)$ outside the sphere $(r>R)$
and $I_2(\x)$ in the sphere $(r<R)$.
These functions obey eq.~(2.1), together with the conditions
$$
I_1=I_2,\quad D_1\n\cdot\nabla I_1=D_2\n\cdot\nabla I_2\quad(r=R),
\eqno(2.22)
$$
expressing respectively the continuity of the intensity
and the conservation of the flux at the boundary of the sphere.
Here $\n$ denotes the outward normal to the sphere.

We look for a solution to eq.~(2.1) which behaves as $I(\x)\approx x$
far from the sphere.
This formally amounts to calculating the dipole moment of a dielectric sphere
in a uniform electric field.
It can be checked that the solution assumes the form
$$
I_1(\x)=\left(1+\frac{P}{r^3}\right)x,\quad I_2(\x)=Ax.
\eqno(2.23)
$$
The outer solution $I_1(\x)$ is the superposition of a uniform gradient
and of a dipole field,
while the inner solution $I_2(\x)$ is a uniform gradient.
The dipole component of $I_1(\x)$ yields the polarizability $P$,
again by eqs.~(2.3), (2.4).

The conditions~(2.22) yield two linear equations for the unknowns
$P$ and $A$, hence
$$
P=\frac{D_1-D_2}{2D_1+D_2}R^3.
\eqno(2.24)
$$
In the present work we are mostly interested in the limit $D_2\to\infty$,
which yields the polarizability of a transparent sphere,
with the same optical index as the medium,
$$
P\dif=-R^3,
\eqno(2.25)
$$
and in the limit $D_2\to 0$,
which yields the polarizability of a totally reflecting sphere,
$$
P\dif=\frac{R^3}{2}.
\eqno(2.26)
$$
Again, these results are expected to hold in the regime $R\gg\ell$.
The dependence of both polarizabilities
on $R/\ell$ will be investigated in sections~3.3 and 3.4.

\vfill\eject
\noindent$\bullet$ {\it The case of a cylinder}
\smallskip

Consider a non-absorbing cylinder of radius $R\gg\ell$, located as above.
Assume again that the diffusion constant reads
$D_1$ in the medium and $D_2$ in the cylinder.
The intensity field can be calculated along the lines
of the above case of a sphere.
We thus obtain
$$
P=\frac{D_1-D_2}{D_1+D_2}R^2.
\eqno(2.27)
$$
Again, the limit $D_2\to\infty$ yields the polarizability of
a transparent cylinder, with the same optical index as the medium,
$$
P\dif=-R^2,
\eqno(2.28)
$$
while the limit $D_2\to 0$ yields the polarizability of a reflecting cylinder,
$$
P\dif=R^2.
\eqno(2.29)
$$
The dependence of both polarizabilities
on $R/\ell$ will be investigated in sections~4.3 and 4.4.

\medskip
\noindent{\bf 3 RADIATIVE TRANSFER THEORY: THE CASE OF A SPHERE}
\smallskip

We turn to the calculation of the capacitance $Q$
or the polarizability $P$ of mesoscopic spheres,
with a radius $R$ comparable to the mean free path $\ell$,
in the three cases listed in the Introduction.
As recalled there, the appropriate framework is RTT,
at least in the regime $\ell\gg\lambda$, corresponding to most experiments.
In the following, we consider for simplicity
isotropic scattering of scalar waves.

Throughout sections~3 and 4,
all lengths are measured in units of the mean free path $\ell$,
for convenience, unless otherwise stated.
So $R$ stands for the dimensionless ratio $R/\ell$, and so on.

\medskip
\noindent{\bf 3.1 Basics of radiative transfer theory}
\smallskip

We first recall the basic concepts of RTT,
following the book by Chandrasekhar~[8].
In a time-independent situation,
the basic quantity is the {\it specific intensity} $I(\x,\n)$,
namely the amount of intensity located at the point $\x$
and propagating into the direction of the unit vector $\n$.
In the case of isotropic scattering,
the local form of the RTT equation reads
$$
\n\cdot\nabla I(\x,\n)=\J(\x)-I(\x,\n),
\eqno(3.1)
$$
in dimensionless units, where the {\it source function} $\J(\x)$ reads
$$
\J(\x)=\int\frac{\d\omega(\n)}{4\pi}I(\x,\n),
\eqno(3.2)
$$
with $\d\omega(\n)$ being the element of solid angle around the direction $\n$.
The source function $\J(\x)$
will be identified with the diffuse intensity $I(\x)$
involved in the diffusion approximation.

The RTT equation~(3.1) admits the formal solution
$$
I(\x,\n)=\int_0^\infty e^{-s}\d s\,\J(\y(\x,\n,s)),
\eqno(3.3)
$$
where $\y(\x,\n,s)$ is the point such that there is a ray
of optical length $s$, starting from point $\y$
and arriving at point $\x$ while pointing into direction $\n$.
We thus obtain the following linear integral form
of the RTT equation for the source function
$$
\J(\x)=\int\frac{\d\omega(\n)}{4\pi}\int_0^\infty e^{-s}\d s\,\J(\y(\x,\n,s)),
\eqno(3.4)
$$
which will be the starting point of all subsequent developments.

In an infinite medium, without any object, we have
$$
\y(\x,\n,s)=\x-\n s,
\eqno(3.5)
$$
so that eq.~(3.4) can be recast as a convolution integral equation of the form
$$
\J(\x)-\int\d^3\y M(\x-\y)\J(\y)=0,
\eqno(3.6)
$$
with a convolution kernel
$$
M(\x-\y)=\int\frac{\d\omega(\n)}{4\pi}\int_0^\infty e^{-s}\d s\,
\delta^{(3)}(\x-\y-\n s)=\frac{e^{-\vert\x-\y\vert}}{4\pi\vert\x-\y\vert^2}.
\eqno(3.7)
$$
Its Fourier transform reads
$$
\widehat M(\q)
=\int M(\x)e^{i\q\cdot\x}\d^3\x
=\int\frac{\d\omega(\n)}{4\pi}\frac{1}{1-i\q\cdot\n}
=\frac{\arctan q}{q}=1-\frac{q^2}{3}+\cdots,
\eqno(3.8)
$$
with $q=\vert\q\vert$.
Therefore the left-hand side of eq.~(3.6)
reads approximately $(-1/3)\nabla^2\J(\x)$, up to higher-order derivatives.
This demonstrates that RTT incorporates the diffusion approximation
in the regime where distances are large with respect
to the mean free path $\ell$,
corresponding to $q\ll 1$ in reduced units.

\medskip
\noindent{\bf 3.2 The capacitance of a totally absorbing (black) sphere}
\smallskip

By definition of a totally absorbing object,
any ray which hits it corresponds to radiation which is absorbed.
A typical such ray, to be discarded from the integral in eq.~(3.4),
is shown as $TX$ in Figure~1.

We introduce spherical co-ordinates $(r,\theta,\varphi)$, such that
$(x,y,z)=(r\cos\theta,$ $r\sin\theta\cos\varphi,$ $r\sin\theta\sin\varphi)$.
Because of the rotational invariance of the problem,
we can choose to locate the observation point $X$ on the positive $x$-axis,
at a distance $r>R$, and the ray in the $x$-$y$ plane,
which coincides with the plane of Figure~1.
The direction $\n$ is marked by the angle $\alpha$ $(0\le\alpha\le\pi)$,
while the incidence angle of the ray at the surface of the sphere
is $\beta$ $(0\le\beta\le\pi/2)$.
Some useful distances and other quantities
related to Figure~1 are listed in Table~1.

We are thus led to split the range of integration over $\n$ and $s$
in eq.~(3.4), and in all similar integrals to appear in the following,
into two domains, respectively called $I$ and $II$,
corresponding to rays which hit the sphere and rays which do not.
Table~1 also gives analytical definitions of these domains,
and of the following shorthand notations
$$
\dint_I\cdots,\qquad\dint_{II}\cdots,\qquad
\dint\full\cdots=\dint_I\cdots+\dint_{II}\cdots
\eqno(3.9)
$$
for the integrals over either domain,
and for the full integral over the whole range of parameters
$(0<\alpha<\pi$ and $0<s<\infty)$.
The latter is normalized:
$$
\dint\full 1=1.
\eqno(3.10)
$$
In the following, it will be advantageous to change variable from $\alpha$
to $\beta$ in integrals over the domain $I$,
and to use the second form of the integration measure given in Table~1.

The RTT equation~(3.4) for a totally absorbing sphere thus reads
$$
\J(\x)=\dint_{II}\J(\x-\n s).
\eqno(3.11)
$$
Along the lines of section~2.3.1,
we look for a rotationally invariant solution $\J(\x)=J(r)$ to eq.~(3.11),
going to unity at large distances.
Eq.~(3.11) for the function $J(r)$ reads
$$
J(r)=\dint_{II}J(\rho),
\eqno(3.12)
$$
with
$$
\rho=\vert\x-\n s\vert=(r^2+s^2-2rs\cos\alpha)^{1/2}.
\eqno(3.13)
$$
We set
$$
J(r)=1-\frac{Q}{r}+F(r),
\eqno(3.14)
$$
where $F(r)$ represents the short-ranged
correction to the solution~(2.3) of the diffusion equation.
This function therefore describes the skin layer around the sphere.
It is expected to fall off exponentially,
with a decay length given by the mean free path,
i.e., $F(r)\sim e^{-r}$ in reduced units.
Some algebra allows to recast eq.~(3.12) as an equation for $F(r)$ itself:
$$
F(r)-\dint\full F(\rho)+\dint_I F(\rho)=S_1(r)+QS_2(r),
\eqno(3.15)
$$
with
$$
S_1(r)=-\dint_I 1,\quad S_2(r)=f(r)+\dint_I\frac{1}{\rho},
\eqno(3.16)
$$
and
$$
f(r)=\frac{1}{r}-\dint\full\frac{1}{\rho}
=\frac{e^{-r}}{r}-\int_r^\infty\frac{e^{-s}\d s}{s}.
\eqno(3.17)
$$

\smallskip\noindent$\bullet$ {\it Small-radius behavior}

The behavior of the capacitance $Q$ for $R\ll 1$ can be derived
by taking the Fourier transform at $\q\to 0$ of eq.~(3.15),
or equivalently its integral over all space.
Indeed, the first two terms of the left-hand side cancel out by this procedure,
because of the behavior~(3.8) of the Fourier-transformed
convolution kernel $\widehat M(\q)$.
The space integral of $f(r)$ is elementary,
while the other contribution to $S_2(r)$ is negligible.
The integral of $S_1(r)$ can be performed exactly,
by changing variables in domain $I$ from $\alpha$ to $\beta$,
and from $R<r<\infty$ to $R\cos\beta<w<\infty$,
according to the definitions given in Table~1.
We thus obtain
$$
\eqalign{
\int_0^\infty r^2\,\d r\,f(r)&=\frac{1}{3},\cr
\int_R^\infty r^2\,\d r\,S_1(r)
&=-\frac{R^2}{2}\int_0^{\pi/2}\sin\beta\cos\beta\,\d\beta
\int_{R\cos\beta}^\infty e^{R\cos\beta-w}\d w=-\frac{R^2}{4}.
}
\eqno(3.18)
$$

By inserting these expressions into eq.~(3.15),
we obtain the following prediction
$$
Q\approx\frac{3R^2}{4\ell},
\eqno(3.19)
$$
in physical units, for $R\ll\ell$.
The leading correction to this result is of relative order $R/\ell$.
It cannot be predicted by elementary means,
since it involves, among other contributions,
the space integral of the third term in the left-hand side of eq.~(3.15).

The power of $R/\ell$ involved in the leading correction
to the estimate~(3.19) can be found more rapidly as follows.
Consider the integral~(3.18) of the source function $S_1(r)$.
The lower bound of this integral has been set to the natural value $r=R$,
since domains $I$ and $II$ are well-defined for $r>R$ only.
Changing the lower bound from $r=R$ to $r=0$, say,
would alter the value of the integral by an amount of relative order $R/\ell$.
By consistency of the whole approach,
the relative correction to eq.~(3.19) is expected to be of the same order.
This {\it rule of thumb} will be used throughout the following.

The leading-order result~(3.19) has been obtained independently in ref.~[7].
Paasschens and~'t~Hooft~[6] obtain the estimate $Q\approx R^2/\ell$,
by evaluating the average dwell time,
i.e., the mean time spent by radiation inside the object.
The discrepancy between their result and eq.~(3.19)
can be explained by noticing that the appropriate quantity to consider
in their approach is rather the inverse of the mean inverse dwell time~[15].

\smallskip\noindent$\bullet$ {\it Large-radius behavior}

If the radius of the sphere is large $(R\gg 1)$,
its surface is flat at the scale of the mean free path,
where $F(r)$ exhibits appreciable variations.
Hence in this limit eq.~(3.12) is expected to become
a RTT equation in a half-space.
This is indeed the case.
Let us set
$$
r=R+\tau\quad(0<\tau<\infty),
\eqno(3.20)
$$
and $J(r)=\Gamma(\tau)$.
We have, to leading order as $R\to\infty$,
$$
\rho=R+\tau',\quad\hbox{with}\quad\tau'=\tau-s\cos\alpha,
\eqno(3.21)
$$
this quantity being positive in domain $II$,
so that eq.~(3.12) can be recast as
$$
\Gamma(\tau)-\int_0^\infty\d\tau'\,M_1(\tau-\tau')\Gamma(\tau')=0,
\eqno(3.22)
$$
where
$$
M_1(\tau)=\dint\full\delta(\tau-s\cos\alpha)
=\int_0^1\frac{\d\mu}{2\mu}e^{-\vert\tau\vert/\mu}
=-\frac{1}{2}{\rm Ei}\big(-\vert\tau\vert\big),
\eqno(3.23)
$$
with ${\rm Ei}(x)$ being the exponential integral function.
The full integral appears in eq.~(3.23), while domain $I$ does not contribute,
because $\alpha_0=\pi/2$ in this regime.

Eq.~(3.22) is nothing but the homogeneous Schwarzschild-Milne
(SM) integral equation of RTT in a half-space with a free boundary.
The function $M_1(x)$ is the associated Milne kernel for isotropic scattering,
with the subscript~1 reminding that $M_1(x)$
is a one-dimensional projection of the RTT kernel $M(\x)$ of eq.~(3.7).
The normalized solution to eq.~(3.22) behaves as~[12--14,~1]
$$
\Gamma(\tau)\approx\tau+\tau_0\quad(\tau\gg 1),
\eqno(3.24)
$$
where
$$
\tau_0
=\frac{1}{\pi}\int_0^{\pi/2}\frac{\d x}{\sin^2x}\ln\frac{\tan^2x}{3(1-x\cot x)}
=0.710446090
\eqno(3.25)
$$
is Milne's extrapolation length~[8,~11].
This term reminds that the extrapolation of the diffusive behavior~(3.24)
vanishes at $\tau=-\tau_0$.

By requiring that expression~(3.24) is proportional
to the solution of the diffusion equation, $J(r)=1-Q/r$,
in the range $1\ll\tau\ll R$ [see eq.~(3.14)], we obtain the estimate
$$
Q\approx\frac{R^2}{R+\tau_0\ell}\approx R\left(1-\frac{\tau_0\ell}{R}\right),
\eqno(3.26)
$$
in physical units, for $R\gg\ell$.
The leading-order behavior of eq.~(3.26) agrees
with the prediction~(2.19) of the diffusion approximation, $Q\dif=R$.
The first correction to this result has a simple interpretation:
the diffusion approximation
has to be corrected by taking into account a skin layer of thickness
$\tau_0\ell$, and requiring that the diffusive field does not
vanish exactly at the surface of the sphere,
but at a distance $\tau_0\ell$ inside the sphere.

For arbitrary values of the radius $R$ of the sphere,
its capacitance $Q$ is such that the inhomogeneous
linear equation~(3.15) for $F(r)$ has a solution which decays rapidly
as $r\to\infty$.
This algorithm can be implemented numerically,
along the lines of section~V of ref.~[12].
The idea is to discretize eq.~(3.15) into an inhomogeneous system of linear
equations, and to solve the latter system numerically,
first with a right-hand side equal to $S_1(r)$, yielding a solution $F_1(r)$,
then with a right-hand side equal to $S_2(r)$, yielding a solution $F_2(r)$,
so that $Q$ is determined by requiring that the linear combination
$F(r)=F_1(r)+QF_2(r)$ decays rapidly as $r\to\infty$.

Figure~2 shows a plot of the size factor $Q/Q\dif$ against $R/\ell$.
The numerical data obtained by the above method smoothly
interpolate between the asymptotic behaviors~(3.19) and~(3.26), listed in
Table~2.

\medskip
\noindent{\bf 3.3 The polarizability of a transparent sphere}
\smallskip

Consider now a transparent sphere,
having the same index of refraction as the medium,
e.g. a cavity containing no scatterers.
The status of the rays in domain $II$, which do not hit the sphere,
is unchanged with respect to the previous case of an absorbing sphere:
$\y(\x,\n,s)$ is again given by eq.~(3.5).
A typical ray in domain $I$, which hits the sphere,
is the ray $TX$ of Figure~1.
The part $BC$ does not contribute to the optical length of the ray,
which reads $s=\vert TB\vert+\vert CX\vert=\vert TB\vert+s_0$.
The point $T$ is therefore located at
$$
\y_T(\x,\n,s)=\x-s_T\n,\quad\hbox{with}\quad
s_T-s=\vert BC\vert=2R\cos\beta=2(R^2-r^2\sin^2\alpha)^{1/2}.
\eqno(3.27)
$$
The RTT equation~(3.4) for a transparent sphere thus reads
$$
\J(\x)=\dint_{I}\J(\x-\n s_T)+\dint_{II}\J(\x-\n s).
\eqno(3.28)
$$

Eq.~(3.28) admits the solution $\J(\x)=1$,
showing that a transparent sphere has a vanishing capacitance,
as anticipated since it does not absorb radiation.
In order to calculate its polarizability $P$,
in analogy with the outer solution $I_1(\x)$ of eq.~(2.23),
we look for a solution to eq.~(3.28) of the form
$$
\J(\x)=xK(r)=rK(r)\cos\theta,
\eqno(3.29)
$$
with $K(r)$ going to unity at large distances.
Eq.~(3.28) then becomes the following equation for the unknown function $K(r)$:
$$
rK(r)=\dint_{I}(r-s_T\cos\alpha)K(\rho_T)+\dint_{II}(r-s\cos\alpha)K(\rho),
\eqno(3.30)
$$
where $\rho$ has been defined in eq.~(3.13), and with
$$
\rho_T=(r^2+s_T^2-2rs_T\cos\alpha)^{1/2}.
\eqno(3.31)
$$
It can be checked that eq.~(3.30) is indeed independent
of the particular choice we made,
namely to put the observation point on the positive $x$-axis.

We now set
$$
K(r)=1+\frac{P}{r^3}+G(r),
\eqno(3.32)
$$
where $G(r)$ again represents the short-ranged skin-layer
correction to the solution~(2.3) of the diffusion equation.
The equation for $G(r)$ reads
$$
rG(r)+\dint\full(s\cos\alpha-r)G(\rho)
+\dint_I\big[(s_T\cos\alpha-r)G(\rho_T)+(r-s\cos\alpha)G(\rho)\big]
=S_1(r)+PS_2(r),
\eqno(3.33)
$$
with
$$
\eqalign{
S_1(r)&=\dint_I(s-s_T)\cos\alpha=-2R\dint_I\cos\alpha\cos\beta,\cr
S_2(r)&=-\frac{e^{-r}}{r^2}+\dint_I
\left(\frac{r-s_T\cos\alpha}{\rho_T^3}+\frac{s\cos\alpha-r}{\rho^3}\right).
}
\eqno(3.34)
$$

\smallskip\noindent$\bullet$ {\it Small-radius behavior}

The behavior of the polarizability $P$ for $R\ll 1$ can be derived
by taking the Fourier transform at $\q\to 0$ of eq.~(3.33),
after multiplication by $\cos\theta$.
The Fourier transform of the function $\J(\x)$ defined in eq.~(3.29) reads
$$
\widehat\J(\q)=\frac{4i\pi}{q^2}\int_0^\infty r\,\d r\,K(r)
(\sin qr-qr\cos qr)
\approx\frac{4i\pi q}{3}\int_0^\infty r^4\,\d r\,K(r)\quad(q\to 0).
\eqno(3.35)
$$
Therefore both sides of eq.~(3.33) have to be integrated
with a weight $r^3\,\d r$.
The integral of $S_1(r)$ can again be estimated
by changing variables from $\alpha$ to $\beta$ and from $r$ to $w$:
$$
\int_R^\infty r^3\,\d r\,S_1(r)
=-R^3\int_0^{\pi/2}\sin\beta\cos^2\beta\,\d\beta
\int_{R\cos\beta}^\infty e^{R\cos\beta-w}w\,\d w
=-R^3\left(\frac{1}{3}+\frac{R}{4}\right).
\eqno(3.36)
$$
The integral of the first contribution to $S_2(r)$ is elementary,
while the second contribution is negligible.

We thus obtain the estimate
$$
P\approx -\frac{R^3}{3}\left(1+\frac{3R}{4\ell}\right),
\eqno(3.37)
$$
in physical units.
The rule of thumb exposed in section~3.2 predicts that the leading
correction to the estimate~(3.37) is of relative order $R^2/\ell^2$.

\smallskip\noindent$\bullet$ {\it Large-radius behavior}

When the radius of the sphere is large $(R\gg 1)$,
eq.~(3.30) again becomes a RTT equation in a half-space.
With the definition~(3.20), we have
$$
\matrix{
\rho=R+\tau',&\hbox{with}&\tau'=\tau-s\cos\alpha>0,
&\hbox{in domain }II,\cr
\rho_T=R+\tau'',&\hbox{with}&\tau''=s\cos\alpha-\tau>0,
&\hbox{in domain }I.\hfill\cr
}
\eqno(3.38)
$$
Setting $K(r)=\Gamma(\tau)$, eq.~(3.30) can be recast as
$$
\Gamma(\tau)
-\int_0^\infty\d\tau'\big[M_1(\tau-\tau')+M_1^L(\tau+\tau')\big]\Gamma(\tau')
=0,
\eqno(3.39)
$$
where
$$
M_1^L(x)=\dint\full(1-2\cos^2\alpha)\,\delta(x-s\cos\alpha)
=\int_0^1\frac{\d\mu}{2\mu}(1-2\mu^2)e^{-\vert x\vert/\mu}.
\eqno(3.40)
$$
We thus obtain the SM integral equation of RTT
in a half-space with partial reflection at the boundary,
characterized by an effective intensity reflection coefficient
$$
R(\mu)=1-2\mu^2.
\eqno(3.41)
$$
The reflection coefficient is negative for $\alpha$ small enough,
namely for $\mu=\cos\alpha>1/\sqrt2$, i.e., $\alpha<\pi/4$.
This observation is less surprising
if we remember that we are only investigating
the contribution to the intensity field of the dipole of the spherical object.
This contribution is not a positive function,
because of its angular dependence, given by eq.~(3.29).

The normalized solution to eq.~(3.39) behaves as
$$
\Gamma(\tau)\approx\tau+\tau_{01}\quad(\tau\gg 1).
\eqno(3.42)
$$
The extrapolation length $\tau_{01}$ cannot be calculated analytically.
We have evaluated it numerically,
again along the lines of section~V of ref.~[12], obtaining thus
$$
\tau_{01}\approx 0.4675.
\eqno(3.43)
$$

By requiring that expression~(3.42) is proportional
to the solution the diffusion equation, $K(r)=1+P/r^3$,
in the range $1\ll\tau\ll R$ [see eq.~(3.32)], we obtain the estimate
$$
P\approx-\frac{R^4}{R+3\tau_{01}\ell}
\approx-R^3\left(1-\frac{3\tau_{01}\ell}{R}\right),
\eqno(3.44)
$$
in physical units, for $R\gg\ell$.
This result agrees, to leading order,
with the prediction~(2.25) of the diffusion approximation, $P\dif=-R^3$.

Figure~3 shows a plot of the size factor $P/P\dif$ against $R/\ell$.
The numerical data interpolate between the asymptotic behaviors~(3.37)
and~(3.44), listed in Table~2.

\medskip
\noindent{\bf 3.4 The polarizability of a reflecting sphere}
\smallskip

Consider now a reflecting sphere.
The status of the rays in domain $II$, which do not hit the sphere,
is unchanged with respect to the two previous cases.
A typical ray in domain $I$, which hits the sphere,
is the ray $RCX$ on Figure~1.
We have $s-s_0=\vert RC\vert$, while $s_0=\vert XC\vert$.
The point $R$ is thus located at
$$
\y_R(\x,\n,s):\left\{\matrix{
x_R=r-s_0\cos\alpha+(s-s_0)\cos(2\beta-\alpha),\cr
y_R=s_0\sin\alpha+(s-s_0)\sin(2\beta-\alpha).\hfill\cr
}\right.
\eqno(3.45)
$$
The RTT equation~(3.4) for a reflecting sphere thus reads
$$
\J(\x)=\dint_{I}\J(\y_R)+\dint_{II}\J(\x-\n s).
\eqno(3.46)
$$

We again look for a solution to eq.~(3.46) of the form
$$
\J(\x)=xK(r)=rK(r)\cos\theta,
\eqno(3.47)
$$
with $K(r)$ going to unity at large distances.
Eq.~(3.46) becomes
$$
rK(r)=\dint_{I}x_RK(\rho_R)+\dint_{II}(r-s\cos\alpha)K(\rho),
\eqno(3.48)
$$
where $\rho$ has been defined in eq.~(3.13), and with
$$
\rho_R=\vert\y_R\vert=(x_R^2+y_R^2)^{1/2}.
\eqno(3.49)
$$
Setting again
$$
K(r)=1+\frac{P}{r^3}+G(r),
\eqno(3.50)
$$
the equation for $G(r)$ reads
$$
rG(r)+\dint\full(s\cos\alpha-r)G(\rho)
+\dint_I\big[(r-s\cos\alpha)G(\rho)-x_R G(\rho_R)\big]=S_1(r)+PS_2(r),
\eqno(3.51)
$$
with
$$
\eqalign{
S_1(r)&=\dint_I(s\cos\alpha-r+x_R)
=2\dint_I(s-s_0)\cos\beta\cos(\alpha-\beta),\cr
S_2(r)&=-\frac{e^{-r}}{r^2}+\dint_I
\left(\frac{x_R}{\rho_R^3}+\frac{s\cos\alpha-r}{\rho^3}\right).
}
\eqno(3.52)
$$

\smallskip
\noindent$\bullet$ {\it Small-radius behavior}

The behavior of the polarizability $P$ for $R\ll 1$ can again be derived
along the lines of the previous section.
The integral
$$
\eqalign{
\int_R^\infty r^3\,\d r\,S_1(r)
&=R^2\int_0^{\pi/2}\sin\beta\cos^2\beta\,\d\beta
\int_{R\cos\beta}^\infty e^{R\cos\beta-w}(w\cos\beta+R\sin^2\beta)\d w\cr
&=R^2\left(\frac{1}{4}+\frac{R}{3}\right)
}
\eqno(3.53)
$$
leads to the estimate
$$
P\approx\frac{R^2\ell}{4},
\eqno(3.54)
$$
in physical units, up to a correction of relative order $R/\ell$.
This result agrees with the leading-order prediction
of Lancaster and Nieuwenhuizen~[7], $P\approx\stot\ell/(4\pi)$,
with the total cross-section $\stot$ assuming
its geometrical value $\stot=\pi R^2$.

\vfill\eject
\noindent$\bullet$ {\it Large-radius behavior}

When the radius of the sphere is large, eq.~(3.48) again becomes
a RTT equation in a half-space.
Indeed, eqs.~(3.38) still hold, to leading order.
Setting $K(r)=\Gamma(\tau)$, eq.~(3.48) can be recast as
$$
\Gamma(\tau)
-\int_0^\infty\d\tau'\big[M_1(\tau-\tau')+M_1(\tau+\tau')\big]\Gamma(\tau')=0.
\eqno(3.55)
$$
We thus obtain the SM integral equation of RTT
in a half-space with total reflection at the boundary.
This equation admits a constant solution,
which we normalize by setting $\Gamma(\tau)=1$.

In order to unravel skin-layer phenomena in the present case,
a careful analysis of corrections to the RTT equation~(3.55) is needed.
Let us anticipate that we have to work up to order $1/R^3$ included, and set
$$
\Gamma(\tau)=1+\frac{\Gamma_1(\tau)}{R}+\frac{\Gamma_2(\tau)}{R^2}
+\frac{\Gamma_3(\tau)}{R^3}+\cdots,
\eqno(3.56)
$$
with the normalization condition $\Gamma(0)=1$,
hence $\Gamma_1(0)=\Gamma_2(0)=\Gamma_3(0)=0$.

The successive correction terms of eq.~(3.56)
obey inhomogeneous SM integral equations, which can be obtained recursively,
by systematically expanding the various quantities
as power series in $1/R$, keeping $\tau$ and $s$ fixed.
For example we have
$$
s_0=\frac{\tau}{\cos\alpha}+\frac{\sin^2\alpha}{2\cos^3\alpha}\frac{\tau^2}{R}
+\frac{\sin^4\alpha}{2\cos^5\alpha}\frac{\tau^3}{R^2}
+\frac{\sin^4\alpha(5-4\cos^2\alpha)}{8\cos^7\alpha}\frac{\tau^4}{R^3}+\cdots,
\eqno(3.57)
$$
and similar series expansions for $\beta$, $x_R$, $\rho_R$, and so on.
Some algebra yields successively
$$
\Gamma_1(\tau)-\int_0^\infty\d\tau'\big[M_1(\tau-\tau')+M_1(\tau+\tau')\big]
\big(\Gamma_1(\tau')+\tau'-\tau\big)=0,
\eqno(3.58)
$$
whose solution, normalized to $\Gamma_1(0)=0$, is
$$
\Gamma_1(\tau)=-\tau,
\eqno(3.59)
$$
and
$$
\Gamma_2(\tau)
-\int_0^\infty\d\tau'\big[M_1(\tau-\tau')+M_1(\tau+\tau')\big]\Gamma_2(\tau')
=-\frac{4}{3},
\eqno(3.60)
$$
whose normalized solution is
$$
\Gamma_2(\tau)=2\tau^2.
\eqno(3.61)
$$
The determination of the third correction term is less easy.
Anticipating the behavior
$\Gamma_3(\tau)=k_1\tau+k_2\tau^2+k_3\tau^3+\gamma_3(\tau)$,
we obtain by consistency $k_3=-10/3$, $k_2=0$, and
$$
\eqalign{
\gamma_3(\tau)
&-\int_0^\infty\d\tau'\big[M_1(\tau-\tau')+M_1(\tau+\tau')\big]\gamma_3(\tau')
\cr
&=\Big((k_1+8)\tau^2-\tau^4\Big)M_1(\tau)
+\frac{1}{2}\Big(k_1+2-(k_1+6)\tau-\tau^2+\tau^3\Big)e^{-\tau},
}
\eqno(3.62)
$$
with $\gamma_3(0)=0$.
A necessary condition for eq.~(3.62) to admit a bounded solution
is that the integral of its right-hand side
over the range $0<\tau<\infty$ vanishes.
This determines $k_1=-4/5$, so that we have
$$
\Gamma_3(\tau)=-\frac{10}{3}\tau^3-\frac{4}{5}\tau+\gamma_3(\tau).
\eqno(3.63)
$$
The calculation of the bounded part $\gamma_3(\tau)$
cannot be performed analytically.

By requiring that expression~(3.56) is proportional
to the solution of the diffusion equation, $K(r)=1+P/r^3$,
in the range $1\ll\tau\ll R$ [see eq.~(3.50)], we obtain the estimate
$$
P\approx\frac{R^3}{2}\left(1+\frac{6\ell^2}{5R^2}\right),
\eqno(3.64)
$$
in physical units, for $R\gg\ell$.
The leading-order behavior of eq.~(3.64) agrees with the prediction
of the diffusion approximation~(2.26), $P\dif=R^3/2$.
The correction term in $\ell^2/R^2$ entirely comes from
the term in $\tau$ in eq.~(3.63),
while eqs.~(3.59),~(3.61), and the leading term in $\tau^3$ in eq.~(3.63)
just reproduce the expansion of the diffusive solution for $\tau\ll R$.

Figure~4 shows a plot of the size factor $P/P\dif$ against $R/\ell$.
The numerical data interpolate between the asymptotic behaviors~(3.54)
and~(3.64), listed in Table~2.

To close up, we mention that the result~(3.64) can be recovered
by the following more direct, albeit less systematic, alternative route.
Assuming the behavior
$$
P\approx\frac{R^3}{2}+AR,
\eqno(3.65)
$$
and systematically using expansions such as eq.~(3.57) for $s_0$,
we can recast the right-hand side of eq.~(3.51) as
$$
S_1(r)+PS_2(r)\approx\frac{1}{R^2}
\int_0^{\pi/2}\frac{\sin\alpha\,\d\alpha}{2}
\int_{\tau\!/\!\cos\alpha}^\infty e^{-s}\d s\,\Phi(\tau,\alpha,s),
\eqno(3.66)
$$
with
$$
\Phi(\tau,\alpha,s)=
4A\cos^2\alpha+(7\cos^2\alpha-3)\tau^2+(9-17\cos^2\alpha)\tau s\cos\alpha
+(10\cos^2\alpha-6)s^2\cos^2\alpha.
\eqno(3.67)
$$
Setting $G(r)=\widetilde\Gamma(\tau)/R^3$,
eq.~(3.51) implies that $\widetilde\Gamma(\tau)$ obeys
an inhomogeneous SM equation, with a right-hand side given by eq.~(3.66),
without the prefactor $1/R^2$.
In order for this SM equation
to have a bounded solution $\widetilde\Gamma(\tau)$,
it is necessary that the $\tau$-integral of its right-hand side vanishes.
We thus obtain after some algebra $A=3/5$, in agreement with eq.~(3.64).
The corresponding solution,
normalized by the condition $\widetilde\Gamma(0)=0$,
reads $\widetilde\Gamma(\tau)=(3/2)\gamma_3(\tau)$.

\vfill\eject
\noindent{\bf 4 RADIATIVE TRANSFER THEORY: THE CASE OF A CYLINDER}
\smallskip

We turn to the calculation of the effective radius $\R$
or the polarizability $P$ of cylinders.
This situation is of special interest,
since it corresponds to the experiments by den Outer et al.~[2].
We recall that throughout this section
lengths are measured in units of the mean free path $\ell$,
unless otherwise stated.

\medskip
\noindent{\bf 4.1 Radiative transfer theory in cylindrical geometry}
\smallskip

We are led to consider RTT in the presence of a cylindrical object
of radius $R$,
whose axis is parallel to the $z$-axis and passes through the origin.
In such a circumstance, the source function $\J(\x)$
is independent of the co-ordinate $z$.
We denote by $\x=\x_\perp=(x,y)$ the position in the plane perpendicular
to the $z$-axis, and we use cylindrical co-ordinates $(r,\theta,z)$,
such that $(x,y)=(r\cos\theta,r\sin\theta)$.
The $x$-$y$ plane again coincides with the plane of Figure~1.

The key ingredient of this analysis
will be the two-dimensional projection $M_2(\s)$ of the
RTT kernel $M(\x)$ of eq.~(3.7),
defined in analogy with the one-dimensional projection $M_1(\tau)$ of
eq.~(3.23), namely
$$
M_2(\s)=\int\frac{\d\omega(\n)}{4\pi}\int_0^\infty e^{-s}\d s\,
\delta(\s-sn_\Vert)=\int_0^{\pi/2}\d\phi\,e^{-\s/\cos\phi},
\eqno(4.1)
$$
where $\phi$ is the angle between the unit vector $\n$ and the $x$-$y$ plane,
with $0\le\phi\le\pi/2$,
so that $n_\Vert=\cos\phi$ is the length of the projection
of $\n$ onto this plane.
The last integral expression of eq.~(4.1) will allow us to calculate
the kernel function $M_2(\s)$ numerically,
and to perform the required analytical calculations involving this function.
We mention for completeness that its derivative is equal to
$\d M_2(\s)/\d\s=-K_0(\s)$, with $K_0$ being the modified Bessel function.

\medskip
\noindent{\bf 4.2 The effective radius of a totally absorbing (black) cylinder}
\smallskip

We first investigate the effective radius $\R$
of an absorbing cylinder of radius $R$.
In analogy with the spherical geometry dealt with in section~3,
we split the range of integration over $\s$ and $\alpha$ into
two domains, $I$ and $II$, respectively
corresponding to rays which hit the cylinder and which do not.
The definitions of these domains,
and of the corresponding integration measures, are given in Table~1.

In analogy with eq.~(3.11), the RTT equation for an absorbing cylinder reads
$$
\J(\x)=\dint_{II}\J(\x-\n\s),
\eqno(4.2)
$$
with $\n$ being the unit vector of the plane
with a direction marked by the angle $\alpha$.
Along the lines of sections~2.3.1 and 3.2,
we look for a cylindrically symmetric solution $\J(\x)=J(r)$ to eq.~(4.2).
This equation becomes
$$
J(r)=\dint_{II}J(\rho),
\eqno(4.3)
$$
with
$$
\rho=\vert\x-\n\s\vert=(r^2+\s^2-2r\s\cos\alpha)^{1/2}.
\eqno(4.4)
$$
We normalize the solution $J(r)$ by setting
$$
J(r)=\ln\frac{r}{\R}+F(r),
\eqno(4.5)
$$
where the function $F(r)$, describing the skin-layer effect,
is again expected to decay exponentially as $r\to\infty$.
We recast eq.~(4.3) as an equation for $F(r)$ itself:
$$
F(r)-\dint\full F(\rho)+\dint_I F(\rho)=S_1(r)+S_2(r)\ln\R,
\eqno(4.6)
$$
with
$$
S_1(r)=f(r)-\dint_I\ln\rho,\quad S_2(r)=\dint_I 1,
\eqno(4.7)
$$
and
$$
f(r)=\dint\full\ln\frac{\rho}{r}=\int_r^\infty M_2(\s)\,\d\s\ln\frac{\s}{r}.
\eqno(4.8)
$$

\smallskip\noindent$\bullet$ {\it Small-radius behavior}

The behavior of the effective radius $\R$ for $R\ll 1$ can again be derived
by taking the Fourier transform at $\q\to 0$ of eq.~(4.6),
or equivalently its integral over all space.
The space integrals of $f(r)$ and $S_2(r)$ can be evaluated by
changing variables in domain $I$ from $\alpha$ to $\beta$,
and from $R<r<\infty$ to $R\cos\beta<w<\infty$,
and by inserting the last integral representation~(4.1) for $M_2(\s)$.
We thus obtain
$$
\eqalign{
\int_0^\infty r\,\d r\,f(r)
&=\frac{1}{4}\int_0^\infty\s^2\,M_2(\s)\,\d\s
=\frac{1}{2}\int_0^{\pi/2}\cos^3\phi\,\d\phi=\frac{1}{3},\cr
\int_R^\infty r\,\d r\,S_2(r)
&=\frac{R}{\pi}\int_0^{\pi/2}\cos\beta\,\d\beta\int_0^{\pi/2}\cos\phi\,\d\phi
\int_{R\cos\beta}^\infty e^{(R\cos\beta-w)/\cos\phi}\d w=\frac{R}{4},
}
\eqno(4.9)
$$
while the second contribution to $S_1(r)$ is negligible.

These exact expressions lead to the estimate
$$
\ln\R\approx-\frac{4}{3R}.
\eqno(4.10)
$$

It turns out that this leading-order estimate has an interesting correction
of relative order $R\ln R$,
originating in the space integral of the last term in the left-hand side
of eq.~(4.6), which can be estimated as follows.
The $R\to 0$ limit $F_0(r)$ of the function $F(r)$ obeys the equation
$$
F_0(r)-\dint\full F_0(\rho)=f(r)-\frac{4}{3\pi r}\int_r^\infty M_2(\s)\,\d\s.
\eqno(4.11)
$$
We observe that the right-hand side
of this equation diverges as $-4/(3\pi r)$ as $r\to 0$.
So does its solution:
$$
F_0(r)\approx-\frac{4}{3\pi r}\quad(r\to 0),
\eqno(4.12)
$$
since the integral in the left-hand side of eq.~(4.11) is less singular.
The space integral of the last term in the left-hand side of eq.~(4.6)
can be estimated by setting $\delta=\s-s_0\approx\s+R\cos\beta-r>0$:
$$
\eqalign{
\int_0^\infty r\,\d r\dint_I F_0(\rho)
&\approx-\frac{4R}{3\pi^2}\int_0^\infty\d r\int_0^\infty M_2(\s)\,\d\s
\int_0^{\pi/2}
\frac{\cos\beta\,\d\beta}{\big(R^2+\delta^2-2R\delta\cos\beta\big)^{1/2}}\cr
&\approx-\frac{4R}{3\pi^2}\int_0^\infty M_2(\s)\,\d\s
\left(\ln\frac{2\s}{R}+\frac{\pi}{2}+1\right)
\approx\frac{4R}{3\pi^2}\left(\ln\frac{R}{4}+\euler-\frac{\pi}{2}\right),
}
\eqno(4.13)
$$
with $\euler$ being Euler's constant.

We thus obtain the more accurate estimate
$$
\ln\frac{\R}{\ell}
\approx-\frac{4\ell}{3R}+\frac{16}{3\pi^2}\ln\frac{R}{\ell}+\ln a,
\eqno(4.14)
$$
i.e.,
$$
\R\approx a\ell\left(\frac{R}{\ell}\right)^{\frad{16}{3\pi^2}}
e^{-\frad{4\ell}{3R}},
\eqno(4.15)
$$
in physical units.
It is worth noticing that the correction in $R\ln R$ derived in eq.~(4.13)
is responsible for the occurrence of the power-law prefactor in the
result~(4.15).
The dimensionless absolute prefactor $a$
receives contributions from several corrections
of relative order $R$ to the above estimates.
A fit of the numerical data shown in Figure~5 leads to $a\approx 1.20$.

The capacitance $Q$ thus reads, from eq.~(2.7),
$$
Q\approx\frac{3R}{4\ell}
+\left(\frac{3}{\pi^2}\ln\frac{R}{\ell}-\frac{9}{16}\ln\frac{r_0}{a\ell}\right)
\frac{R^2}{\ell^2}.
\eqno(4.16)
$$
The leading behavior of the capacitance is therefore an intrinsic
characteristic of the absorbing cylinder, while the length scale $r_0$,
involving the geometry of the sample,
only appears logarithmically in the correction term.

\smallskip\noindent$\bullet$ {\it Large-radius behavior}

If the radius of the cylinder is large $(R\gg 1)$,
eq.~(4.3) again becomes the SM equation~(3.22)
of RTT in a half-space with a free boundary.
By requiring that the asymptotic behavior~(3.24) of its solution
is proportional to the solution of the diffusion equation,
$J(r)=\ln(r/\R)$,
in the range $1\ll\tau\ll R$ [see eq.~(4.5)], we obtain the estimate
$$
\R\approx Re^{-\tau_0\ell/R}\approx R\left(1-\frac{\tau_0\ell}{R}\right),
\eqno(4.17)
$$
in physical units, for $R\gg\ell$,
where $\tau_0$ is again Milne's extrapolation length~(3.25).
This result agrees, to leading order,
with the prediction~(2.21) of the diffusion approximation, $\R\dif=R$.

Figure~5 shows a logarithmic plot of the size factor $\R/\R\dif$ against
$R/\ell$.
The numerical data interpolate between the asymptotic behaviors~(4.15)
and~(4.17), listed in Table~2.

\medskip
\noindent{\bf 4.3 The polarizability of a transparent cylinder}
\smallskip

In analogy with eq.~(3.28), the RTT equation for a transparent cylinder reads
$$
\J(\x)=\dint_{I}\J(\x-\n\s_T)+\dint_{II}\J(\x-\n\s),
\eqno(4.18)
$$
with
$$
\s_T=\s+2R\cos\beta=\s+2(R^2-r^2\sin^2\alpha)^{1/2}.
\eqno(4.19)
$$

In order to calculate the polarizability $P$,
along the lines of sections~2.3.2 and 3.3,
we look for a solution to eq.~(4.18) of the form
$$
\J(\x)=xK(r)=rK(r)\cos\theta,
\eqno(4.20)
$$
with $K(r)$ going to unity at large distances.
Eq.~(4.18) becomes
$$
rK(r)=\dint_{I}(r-\s_T\cos\alpha)K(\rho_T)+\dint_{II}(r-\s\cos\alpha)K(\rho),
\eqno(4.21)
$$
where $\rho$ has been defined in eq.~(4.4), and with
$$
\rho_T=(r^2+\s_T^2-2r\s_T\cos\alpha)^{1/2}.
\eqno(4.22)
$$
We set
$$
K(r)=1+\frac{P}{r^2}+G(r),
\eqno(4.23)
$$
where $G(r)$ again represents the short-ranged skin-layer correction
to the solution of the diffusion equation.
The equation for $G(r)$ reads
$$
rG(r)+\dint\full(\s\cos\alpha-r)G(\rho)
+\dint_I\big[(\s_T\cos\alpha-r)G(\rho_T)+(r-\s\cos\alpha)G(\rho)\big]
\!=\!S_1(r)+PS_2(r),
\eqno(4.24)
$$
with
$$
\eqalign{
S_1(r)&=\dint_I(\s-\s_T)\cos\alpha=-2R\dint_I\cos\alpha\cos\beta,\cr
S_2(r)&=-g(r)+\dint_I
\left(\frac{r-\s_T\cos\alpha}{\rho_T^2}+\frac{\s\cos\alpha-r}{\rho^2}\right),
}
\eqno(4.25)
$$
and
$$
g(r)=\frac{1}{r}-\dint\full\frac{r-\s\cos\alpha}{\rho^2}
=\frac{1}{r}\int_r^\infty M_2(\s)\,\d\s.
\eqno(4.26)
$$

\smallskip\noindent$\bullet$ {\it Small-radius behavior}

The behavior of the polarizability $P$ for $R\ll 1$ can again be derived
by taking the Fourier transform at $\q\to 0$ of eq.~(4.24),
after multiplication by $\cos\theta$.
The Fourier transform of the function $\J(\x)$ defined in eq.~(4.20) reads
$$
\widehat\J(\q)=2i\pi\int_0^\infty r^2\,\d r\,K(r)J_1(qr)
\approx i\pi q\int_0^\infty r^3\,\d r\,K(r)\quad(q\to 0),
\eqno(4.27)
$$
where $J_1$ is the Bessel function.
Therefore both sides of eq.~(4.24) have to be integrated with a weight
$r^2\,\d r$.
The integral of $S_1(r)$ can again be estimated
by changing variables from $\alpha$ to $\beta$ and from $r$ to $w$.
We thus obtain
$$
\eqalign{
\int_R^\infty r^2\,\d r\,S_1(r)
&=-\frac{2R^2}{\pi}\int_0^{\pi/2}\cos^2\beta\,\d\beta
\int_0^{\pi/2}\cos\phi\,\d\phi
\int_{R\cos\beta}^\infty e^{(R\cos\beta-w)/\cos\phi}w\,\d w\cr
&=-\frac{R^2}{3}(1+R),\cr
\int_0^\infty r^2\,\d r\,g(r)
&=\frac{1}{2}\int_0^\infty\s^2\,M_2(\s)\,\d\s
=\int_0^{\pi/2}\cos^3\phi\,\d\phi=\frac{2}{3},\cr
}
\eqno(4.28)
$$
while the second contribution to $S_2(r)$ is negligible.
This leads to the estimate
$$
P\approx-\frac{R^2}{2}\left(1+\frac{R}{\ell}\right),
\eqno(4.29)
$$
in physical units, up to a correction of relative order $R^2/\ell^2$.

\smallskip\noindent$\bullet$ {\it Large-radius behavior}

When the radius of the sphere is large $(R\gg 1)$,
eq.~(4.21) again becomes a RTT equation in a half-space.
With the definitions~(3.20), (3.38), and setting again $K(r)=\Gamma(\tau)$,
we obtain
$$
\Gamma(\tau)
-\int_0^\infty\d\tau'\big[M_1(\tau-\tau')+M_1^L(\tau+\tau')\big]\Gamma(\tau')
=0,
\eqno(4.30)
$$
where
$$
\eqalign{
M_1(x)&=\dint\full\delta(x-\s\cos\alpha)
=\int_0^1\frac{\d\mu}{2\mu}e^{-\vert x\vert/\mu},\cr
M_1^L(x)&=\dint\full(1-2\cos^2\alpha)\,\delta(x-\s\cos\alpha)
=\int_0^1\frac{\d\mu}{2\mu}(1-2\mu)e^{-\vert x\vert/\mu}.
}
\eqno(4.31)
$$
We thus obtain the SM equation
in a half-space with partial reflection at the boundary,
characterized by a reflection coefficient
$$
R(\mu)=1-2\mu,
\eqno(4.32)
$$
which is again negative for $\alpha$ small enough,
namely $\mu=\cos\alpha>1/2$, i.e., $\alpha<\pi/3$.

The normalized solution to eq.~(4.30) again behaves as
$$
\Gamma(\tau)\approx\tau+\tau_{01}\quad(\tau\gg 1).
\eqno(4.33)
$$
We have evaluated $\tau_{01}$ numerically,
again along the lines of section~V of ref.~[12], obtaining thus
$$
\tau_{01}\approx 0.2137.
\eqno(4.34)
$$

By requiring that expression~(4.33) is proportional
to the solution of the diffusion equation, $K(r)=1+P/r^2$,
in the range $1\ll\tau\ll R$ [see eq.~(4.23)], we obtain the estimate
$$
P\approx-\frac{R^3}{R+2\tau_{01}\ell}
\approx-R^2\left(1-\frac{2\tau_{01}\ell}{R}\right),
\eqno(4.35)
$$
in physical units, for $R\gg\ell$.
This result agrees, to leading order,
with the prediction~(2.28) of the diffusion approximation, $P\dif=-R^2$.

Figure~6 shows a plot of the size factor $P/P\dif$ against $R/\ell$.
The numerical data interpolate between the asymptotic behaviors~(4.29)
and~(4.35), listed in Table~2.

\medskip
\noindent{\bf 4.4 The polarizability of a reflecting cylinder}
\smallskip

In analogy with eq.~(3.46), the RTT equation for a reflecting cylinder reads
$$
\J(\x)=\dint_{I}\J(\y_R)+\dint_{II}\J(\x-\n\s),
\eqno(4.36)
$$
with
$$
\y_R(\x,\n,s):\left\{\matrix{
x_R=r-s_0\cos\alpha+(\s-s_0)\cos(2\beta-\alpha),\cr
y_R=s_0\sin\alpha+(\s-s_0)\sin(2\beta-\alpha).\hfill\cr
}\right.
\eqno(4.37)
$$
We again look for a solution to eq.~(4.36) of the form
$$
\J(\x)=xK(r)=rK(r)\cos\theta,
\eqno(4.38)
$$
with $K(r)$ going to unity at large distances.
Eq.~(4.36) becomes
$$
rK(r)=\dint_{I}x_RK(\rho_R)+\dint_{II}(r-\s\cos\alpha)K(\rho),
\eqno(4.39)
$$
with the notation~(3.49).
Setting
$$
K(r)=1+\frac{P}{r^2}+G(r),
\eqno(4.40)
$$
the equation for $G(r)$ reads
$$
rG(r)+\dint\full(\s\cos\alpha-r)G(\rho)
+\dint_I\big[(r-\s\cos\alpha)G(\rho)-x_R G(\rho_R)\big]=S_1(r)+PS_2(r),
\eqno(4.41)
$$
with
$$
\eqalign{
S_1(r)&=2\dint_I(\s-s_0)\cos\beta\cos(\alpha-\beta),\cr
S_2(r)&=-g(r)
+\dint_I\left(\frac{x_R}{\rho_R^2}+\frac{\s\cos\alpha-r}{\rho^2}\right),
}
\eqno(4.42)
$$
where $g(r)$ has been defined in eq.~(4.26).

\smallskip\noindent$\bullet$ {\it Small-radius behavior}

The behavior of the polarizability $P$ for $R\ll 1$ can again be derived
along the lines of the previous sections.
We have
$$
\eqalign{
\int_R^\infty r^2\,\d r\,S_1(r)
&=\frac{2R}{\pi}\int_0^{\pi/2}\cos^2\beta\,\d\beta
\int_0^{\pi/2}\cos^2\phi\,\d\phi\cr
&{\hskip 2cm}\times\int_{R\cos\beta}^\infty e^{(R\cos\beta-w)/\cos\phi}
(w\cos\beta+R\sin^2\beta)\d w\cr
&=R\left(\frac{1}{4}+\frac{R}{3}\right),
}
\eqno(4.43)
$$
hence the estimate
$$
P\approx\frac{3R\ell}{8},
\eqno(4.44)
$$
in physical units, up to a correction of relative order $R/\ell$.

\smallskip\noindent$\bullet$ {\it Large-radius behavior}

In analogy with section~3.5, we have been led to study
the solution to the RTT equation,
including corrections of order $1/R^3$ included.
Setting
$$
\Gamma(\tau)=1+\frac{\Gamma_1(\tau)}{R}+\frac{\Gamma_2(\tau)}{R^2}
+\frac{\Gamma_3(\tau)}{R^3}+\cdots,
\eqno(4.45)
$$
we obtain successively
$$
\Gamma_1(\tau)=-\tau,\quad
\Gamma_2(\tau)=\frac{3}{2}\tau^2,\quad
\Gamma_3(\tau)=-2\tau^3-\frac{2}{5}\tau+\gamma_3(\tau).
\eqno(4.46)
$$

By requiring that expression~(4.45) is proportional
to the solution of the diffusion equation, $K(r)=1+P/r^2$,
in the range $1\ll\tau\ll R$ [see eq.~(4.40)], we obtain the estimate
$$
P\approx R^2\left(1+\frac{4\ell^2}{5R^2}\right),
\eqno(4.47)
$$
in physical units, for $R\gg\ell$.
This result agrees, to leading order,
with the prediction~(2.29) of the diffusion approximation, $P\dif=R^2$.

Figure~7 shows a plot of the size factor $P/P\dif$ against $R/\ell$.
The numerical data interpolate between the asymptotic behaviors~(4.44)
and~(4.47), listed in Table~2.

We have also checked that the alternative approach sketched at the end
of section~3.5 permits to recover the result~(4.47).

\medskip
\noindent{\bf 5 DISCUSSION}
\smallskip

In this paper, we have first recalled that the long-range effects
of an isolated embedded object on the diffusive propagation of radiation
in an otherwise homogeneous, non-absorbing turbid medium
can be described within the diffusion approximation~[2,~4--7].
The leading term of a multipole expansion involve
a (negative) charge $q$ (for an absorbing object)
or a dipole moment $\p$ (for a non-absorbing object).
We have emphasized the role of the associated intrinsic characteristics
of a spherical object, namely its capacitance $Q$ and its polarizability $P$.
For cylinders, we have underlined that the right intrinsic quantity
to be considered is its effective radius $\R$.
The intrinsic quantities $Q$, $\R$, $P$
are related to the observables $q$ and $\p$,
which depend on the geometry of the sample, through eqs.~(2.4), (2.7).
We have shown that RTT provides the adequate framework to evaluate
the characteristics $Q$, $\R$, and $P$ of mesoscopic objects,
with a radius $R$ comparable to the mean free path $\ell$,
the latter being much larger than the wavelength $\lambda$.
Considering for simplicity the case of isotropic scattering of scalar waves,
we have worked out this program for spheres and cylinders
of radius $R$, in several situations:

\noindent(i) totally absorbing (black) object,
{\parskip 0pt

\noindent(ii) transparent object, with the same optical index as the medium,

\noindent(iii) totally reflecting object.
}

For large objects $(R\gg\ell)$,
we recover the predictions of the diffusion approximation~[2,~6],
and we give the analytical form of the first correction to this approximation,
which is either proportional to $\ell/R$ or to $(\ell/R)^2$.
In the opposite regime of small objects $(R\ll\ell)$,
our results have some overlap with those obtained by the partial-wave expansion
of Lancaster and Nieuwenhuizen~[7].
In the intermediate situation, i.e., for generic values of the ratio $R/\ell$,
the characteristics $Q$, $\R$, or $P$ differ from the predictions
$Q\dif$, $\R\dif$, or $P\dif$ of the diffusion approximation
by a multiplicative size factor, which depends continuously on $R/\ell$,
and smoothly interpolates between the analytical results
pertaining to both limiting regimes.
For cases~(i) and (ii),
the size factor is a monotonically increasing function of $R/\ell$,
converging to unity from below, with a correction of order $\ell/R$.
For case~(iii), the size factor is a decreasing function of $R/\ell$,
converging to unity from above, with a correction of order $(\ell/R)^2$.

The characteristics $Q$, $\R$, and $P$ of spheres and cylinders are expected
not to be fully universal,
but rather to weakly depend on microscopic details of the scattering process.
The associated size factors have been calculated in this work
for isotropic scattering of scalar waves.
They would be slightly different in other situations,
such as anisotropic scattering of scalar waves,
or Rayleigh scattering of vector waves.
This weak dependence can be illustrated on the example
of large absorbing objects.
Indeed, for both spheres and cylinders, the first correction
to the diffusion approximation has been shown
to be proportional to the extrapolation length $\tau_0$
of the problem of RTT in a half-space with a free boundary.
This quantity is known to be $\tau_0\approx 0.718211$
for very anisotropic scattering of scalar waves~[13],
in units of the scattering mean free path $\ell^\star$,
and $\tau_0\approx 0.712110$
for Rayleigh scattering of electromagnetic waves~[14],
these numbers being respectively $1.09\%$ and $0.23\%$
above the Milne value~(3.25), $\tau_0\approx 0.710446$,
corresponding to isotropic scattering of scalar waves.

Finally, in all the situations considered in this work,
the effect of the embedded object
on a light ray can be described in simple terms.
In cases~(i) and (iii) radiation does not enter the object at all,
while in case~(ii) it propagates ballistically inside the object.
The present RTT approach can be generalized, at least in principle,
to mesoscopic spheres or cylinders with yet other kinds of optical properties.
For instance, situations where radiation is partly absorbed
in the embedded object and/or in the medium,
can be dealt with rather simply, in the general case where
the corresponding absorption lengths are comparable to the size of the object.
A more complex situation is e.g. that of a transparent object,
with an optical index different from that of the medium,
so that light undergoes multiple reflections inside the object.
In this case the RTT equation receives an infinite series of contributions
from the multiply reflected light rays,
weighted by the appropriate extinction factors.
Furthermore, as opposed to the cases treated in the present paper,
the light rays are not contained in a plane in general,
so that the relevant integral equations keep two angular variables.

\medskip
\noindent{\bf Acknowledgments}

Th.M.N. acknowledges useful discussions with
G. Maret, R. Maynard, M. Nieto-Vesperinas, and J. Ricka.

This work has been supported by the HCM network entitled
Transport of Light in Strongly Scattering Media,
under contract no. ERBCHRXCT 930373.

\vfill\eject
{\parindent 0pt
{\bf REFERENCES}
\bigskip

[1] M.C.W. van Rossum and Th.M. Nieuwenhuizen, Rev. Mod. Phys. (to appear
in January 1999).

[2] P.N. den Outer, Th.M. Nieuwenhuizen, and A. Lagendijk, J. Opt. Soc. Am. A
{\bf 10}, 1209 (1993).

[3] J.C. Schotland, J.C. Haselgrove, and J.S. Leigh, Appl. Opt. {\bf 32}, 448
(1993).

[4] S. Feng, F.A. Zeng, and B. Chance, Appl. Opt. {\bf 34}, 3826 (1995).

[5] X.D. Zhu, S.P. Wei, S. Feng, and B. Chance, J. Opt. Soc. Am. A {\bf 13},
494 (1996).

[6] J.C.J. Paasschens and G.W. 't~Hooft, J. Opt. Soc. Am. A {\bf 15}, 1797
(1998); J.C.J. Paasschens, Ph.D. thesis (Leiden, 1997).

[7] D. Lancaster and Th.M. Nieuwenhuizen, Physica A {\bf 256}, 417 (1998).

[8] S. Chandrasekhar, {\it Radiative Transfer} (Dover, New-York, 1960).

[9] V.V. Sobolev, {\it A Treatise on Radiative Transfer} (Van Nostrand,
Princeton, N.J., 1963).

[10] A. Ishimaru, {\it Wave Propagation and Scattering in Random Media}, in 2
volumes (Academic, New-York, 1978).

[11] H.C. van de Hulst, {\it Multiple Light Scattering}, in 2 volumes
(Academic, New-York, 1980).

[12] Th.M. Nieuwenhuizen and J.M. Luck, Phys. Rev. E {\bf 48}, 569 (1993).

[13] E. Amic, J.M. Luck, and Th.M. Nieuwenhuizen, J. Phys. A {\bf 29}, 4915
(1996).

[14] E. Amic, J.M. Luck, and Th.M. Nieuwenhuizen, J. Phys. I France {\bf 7},
445 (1997).

[15] J.C.J. Paasschens, private discussion.

\vfill\eject

{\bf CAPTIONS OF FIGURES AND TABLES}
\bigskip

{\bf Figure 1:}
Geometrical characteristics of rays which hit the object under study
(a sphere centered at $O$ in section~3,
or a cylinder whose axis is parallel to the $z$-axis
and passes through $O$ in section~4).
Table~1 contains the expressions of various quantities attached to this Figure.

{\bf Figure 2:}
Plot of the size factor $Q/Q\dif$ of the capacitance of an absorbing sphere,
against the ratio $R/\ell$.
Full line: outcome of the numerical analysis explained in the text.
Dashed lines: small-radius and large-radius behaviors, as listed in Table~2.

{\bf Figure 3:}
Same as Figure 2, for the size factor $P/P\dif$ of the polarizability
of a transparent sphere.

{\bf Figure 4:}
Same as Figure 2, for the size factor $P/P\dif$ of the polarizability
of a reflecting sphere.

{\bf Figure 5:}
Same as Figure 2, for the size factor $\R/\R\dif$ of the effective radius
of an absorbing cylinder (logarithmic scale).

{\bf Figure 6:}
Same as Figure 2, for the size factor $P/P\dif$ of the polarizability
of a transparent cylinder.

{\bf Figure 7:}
Same as Figure 2, for the size factor $P/P\dif$ of the polarizability
of a reflecting cylinder.

\bigskip
{\bf Table 1:}
Summary of useful definitions, including
distances and other quantities attached to the rays constructed in Figure~1,
domains $I$ and $II$ in both spherical and cylindrical geometries,
and the associated integration measures.

{\bf Table 2:}
Asymptotic behavior at small and large radius
of the various characteristics of embedded spherical
and cylindrical objects studied in the text.
Absorbing objects are characterized by their capacitance $Q$
or their effective radius $\R$,
while non-absorbing, either transparent or reflecting,
objects are characterized by their polarizability $P$.
The large-radius behaviors agree, to leading order,
with the predictions of the diffusion approximation, recalled in section~2.3.
}
\vfill\eject
\centerline{\bf Table 1}
\medskip
$$
\vbox{\init\halign to 16truecm
{\strut#&\vrule#\tabskip=1em plus 2em&
\hfil$#$\hfil&\vrule#&
\hfil$#$\hfil&\vrule#\tabskip 0pt\crr
&&\ &&\ &\cr
&&\hbox{Geometric quantities}
&&\matrix{
r=\vert OX\vert\hfill\cr
w=\vert AX\vert=r\cos\alpha=(r^2-R^2\sin^2\beta)^{1/2}\hfill\cr
\sp\cr
\vert OA\vert=r\sin\alpha=R\sin\beta\hfill\cr
\vert AB\vert=\vert AC\vert=R\cos\beta=(R^2-r^2\sin^2\alpha)^{1/2}\hfill\cr
\sp\cr
\sin\alpha_0=R/r\hfill\cr
s_0=\vert XC\vert=w-R\cos\beta\hfill\cr
\sp\cr
r>R,\sp s_0>0\hfill\cr
}
&\cr
&&\ &&\ &\crr
&&\ &&\ &\cr
&&\matrix{\hbox{Spherical geometry}\cr\hbox{(Section 3)}}
&&\matrix{
I:{\hskip 4.6mm}0<\alpha<\alpha_0\sp\hbox{and}\sp s_0<s<\infty\hfill\cr
\sp\cr
II:\left\{\matrix{
0<\alpha<\alpha_0\ \hbox{and}\ 0<s<s_0\cr
\hbox{or}\cr
\alpha_0<\alpha<\pi\ \hbox{and}\ 0<s<\infty\cr
}\right.
\hfill\cr
\sp\cr\sp\cr
{\displaystyle\dint_{I,II,{\rm full}}\cdots
=\dint_{I,II,{\rm full}}\frac{\sin\alpha\,\d\alpha}{2}e^{-s}\d s\cdots}
\hfill\cr
\sp\cr\sp\cr
{\displaystyle\dint_I\cdots=\frac{R^2}{2r^2}
\int_0^{\pi/2}\sin\beta\cos\beta\,\d\beta
\int_{s_0}^\infty\frac{e^{-s}\d s}{\cos\alpha}\cdots}
\hfill\cr
}
&\cr
&&\ &&\ &\crr
&&\ &&\ &\cr
&&\matrix{\hbox{Cylindrical geometry}\cr\hbox{(Section 4)}}
&&\matrix{
I:{\hskip 4.6mm}0<\alpha<\alpha_0\sp\hbox{and}\sp s_0<\s<\infty\hfill\cr
\sp\cr
II:\left\{\matrix{
0<\alpha<\alpha_0\ \hbox{and}\ 0<\s<s_0\cr
\hbox{or}\cr
\alpha_0<\alpha<\pi\ \hbox{and}\ 0<\s<\infty\cr
}\right.
\hfill\cr
\sp\cr\sp\cr
{\displaystyle\dint_{I,II,{\rm full}}\cdots
=\dint_{I,II,{\rm full}}\frac{\d\alpha}{\pi}M_2(\s)\,\d\s\cdots}
\hfill\cr
\sp\cr\sp\cr
{\displaystyle\dint_I\cdots=\frac{R}{\pi r}\int_0^{\pi/2}\cos\beta\,\d\beta
\int_{s_0}^\infty\frac{M_2(\s)\,\d\s}{\cos\alpha}\cdots}
\hfill\cr
}
&\cr
&&\ &&\ &\crr
}}
$$
\vfill\eject
\centerline{\bf Table 2}
\medskip
$$
\vbox{\init\halign to 16truecm
{\strut#&\vrule#\tabskip=1em plus 2em&
\hfil$#$\hfil&\vrule#&
\hfil$#$\hfil&\vrule#&
\hfil$#$\hfil&\vrule#\tabskip 0pt\crr
&&\ &&\ &&\ &\cr
&&\hbox{Object}
&&\matrix{\hbox{Small-radius behavior}\cr(R\ll\ell)}
&&\matrix{\hbox{Large-radius behavior}\cr(R\gg\ell)}
&\cr
&&\ &&\ &&\ &\crr
&&\ &&\ &&\ &\cr
&&\matrix{\hbox{absorbing sphere}\cr\hbox{(Sec.~3.2, Fig.~2)}}
&&Q\approx\frad{3R^2}{4\ell}
&&\matrix{Q\approx\underbrace{R}_{\displaystyle Q\dif}
\left(1-\frad{\tau_0\ell}{R}\right)\cr
\sp\cr\tau_0\approx0.710446}
&\cr
&&\ &&\ &&\ &\crr
&&\ &&\ &&\ &\cr
&&\matrix{\hbox{transparent sphere}\cr\hbox{(Sec.~3.3, Fig.~3)}}
&&P\approx-\frad{R^3}{3}\left(1+\frad{3R}{4\ell}\right)
&&\matrix{P\approx\underbrace{-R^3}_{\displaystyle P\dif}
\left(1-\frad{3\tau_{01}\ell}{R}\right)\cr
\sp\cr\tau_{01}\approx0.4675}
&\cr
&&\ &&\ &&\ &\crr
&&\ &&\ &&\ &\cr
&&\matrix{\hbox{reflecting sphere}\cr\hbox{(Sec.~3.4, Fig.~4)}}
&&P\approx\frad{R^2\ell}{4}
&&P\approx\underbrace{\frad{R^3}{2}}_{\displaystyle P\dif}
\left(1+\frad{6\ell^2}{5R^2}\right)
&\cr
&&\ &&\ &&\ &\crr
&&\ &&\ &&\ &\cr
&&\matrix{\hbox{absorbing cylinder}\cr\hbox{(Sec.~4.2, Fig.~5)}}
&&\matrix{\R\approx
a\ell\left(\frad{R}{\ell}\right)^{\frad{16}{3\pi^2}}
e^{-\frad{4\ell}{3R}}\cr
\sp\cr a\approx1.20}
&&\matrix{\R\approx\underbrace{R}_{\displaystyle \R\dif}
\left(1-\frad{\tau_0\ell}{R}\right)\cr
\sp\cr\tau_0\approx0.710446}
&\cr
&&\ &&\ &&\ &\crr
&&\ &&\ &&\ &\cr
&&\matrix{\hbox{transparent cylinder}\cr\hbox{(Sec.~4.3, Fig.~6)}}
&&P\approx-\frad{R^2}{2}\left(1+\frad{R}{\ell}\right)
&&\matrix{P\approx\underbrace{-R^2}_{\displaystyle P\dif}
\left(1-\frad{2\tau_{01}\ell}{R}\right)\cr
\sp\cr\tau_{01}\approx0.2137}
&\cr
&&\ &&\ &&\ &\crr
&&\ &&\ &&\ &\cr
&&\matrix{\hbox{reflecting cylinder}\cr\hbox{(Sec.~4.4, Fig.~7)}}
&&P\approx\frad{3R\ell}{8}
&&P\approx\underbrace{R^2}_{\displaystyle P\dif}
\left(1+\frad{4\ell^2}{5R^2}\right)
&\cr
&&\ &&\ &&\ &\crr
}}
$$
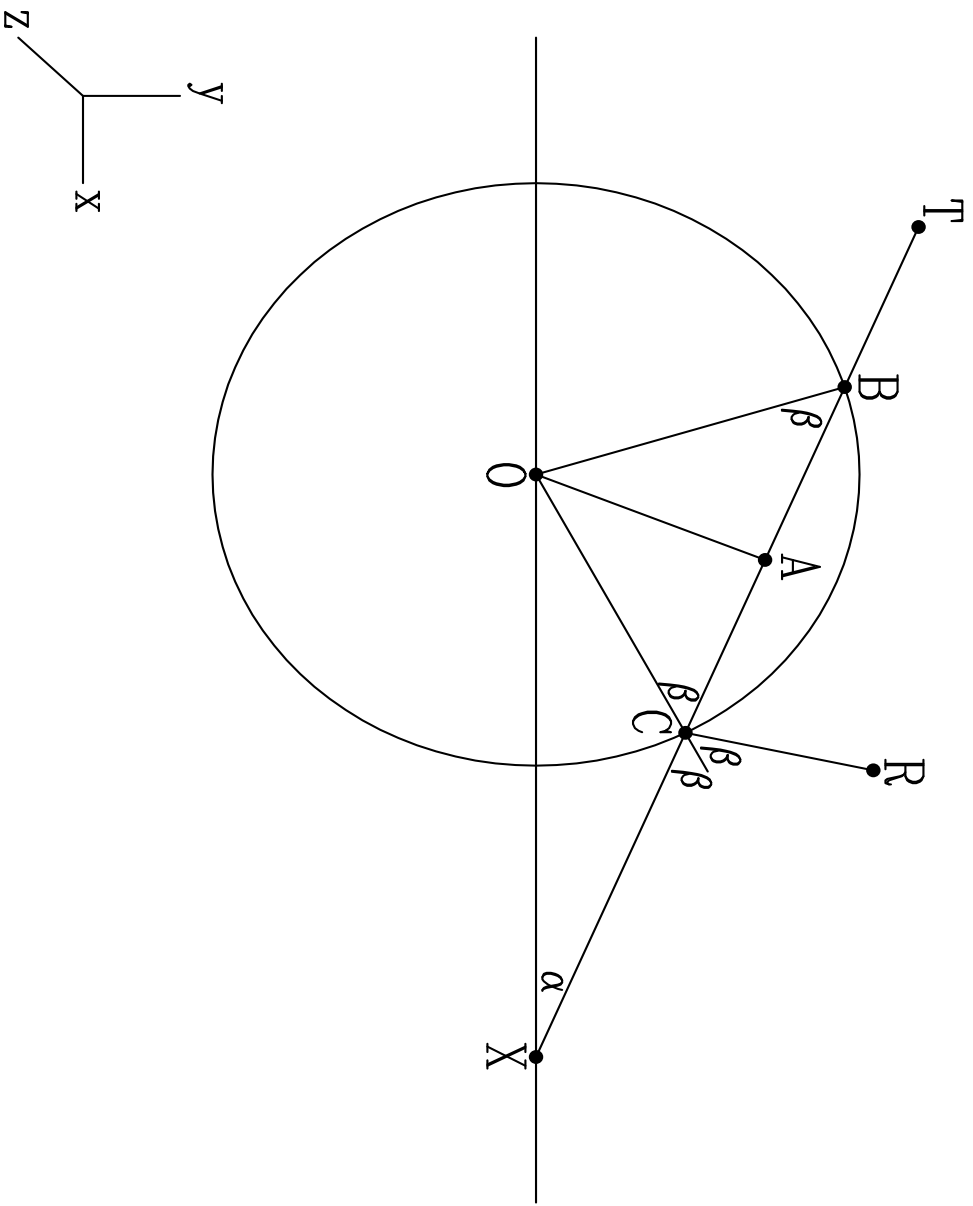
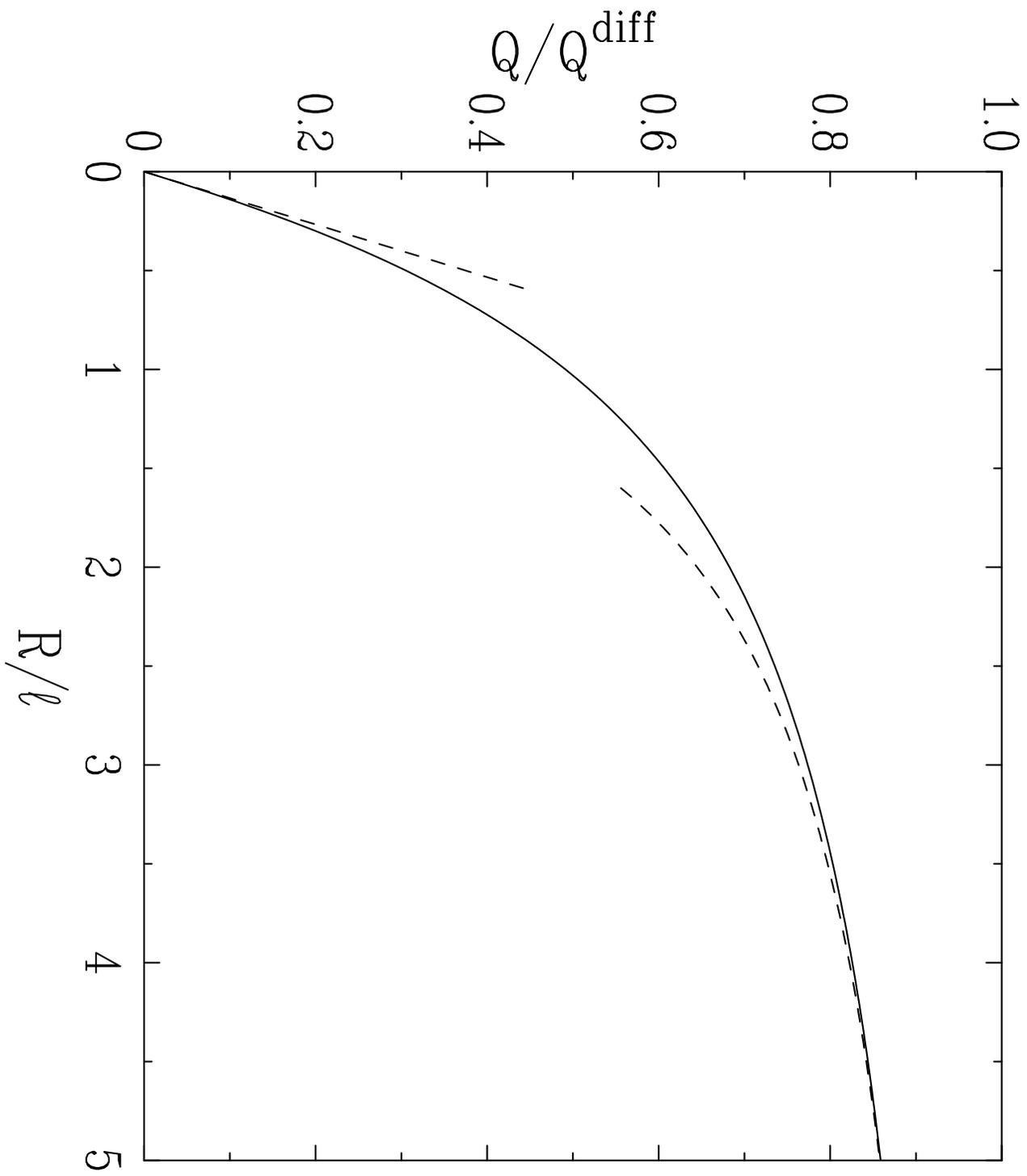
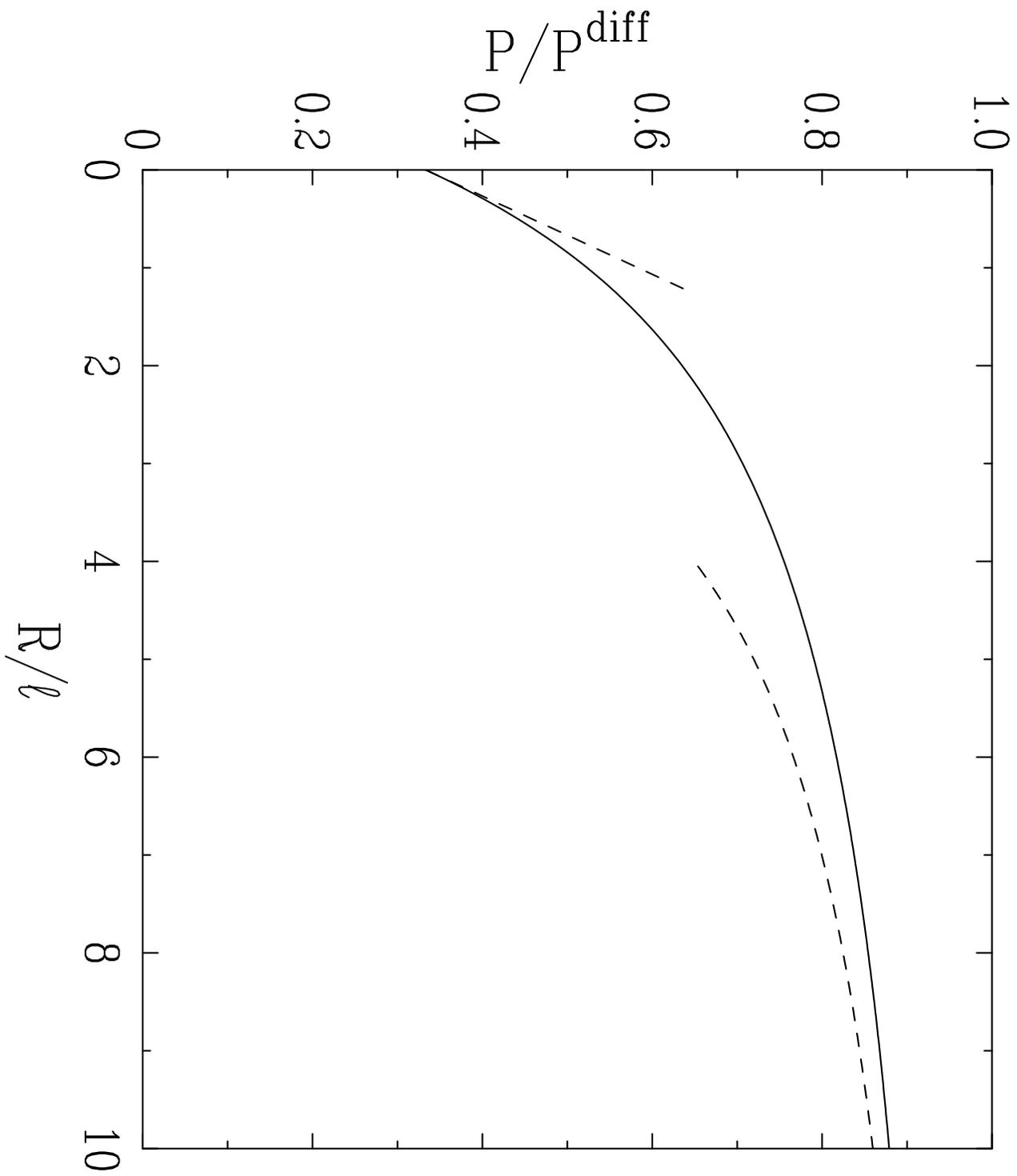
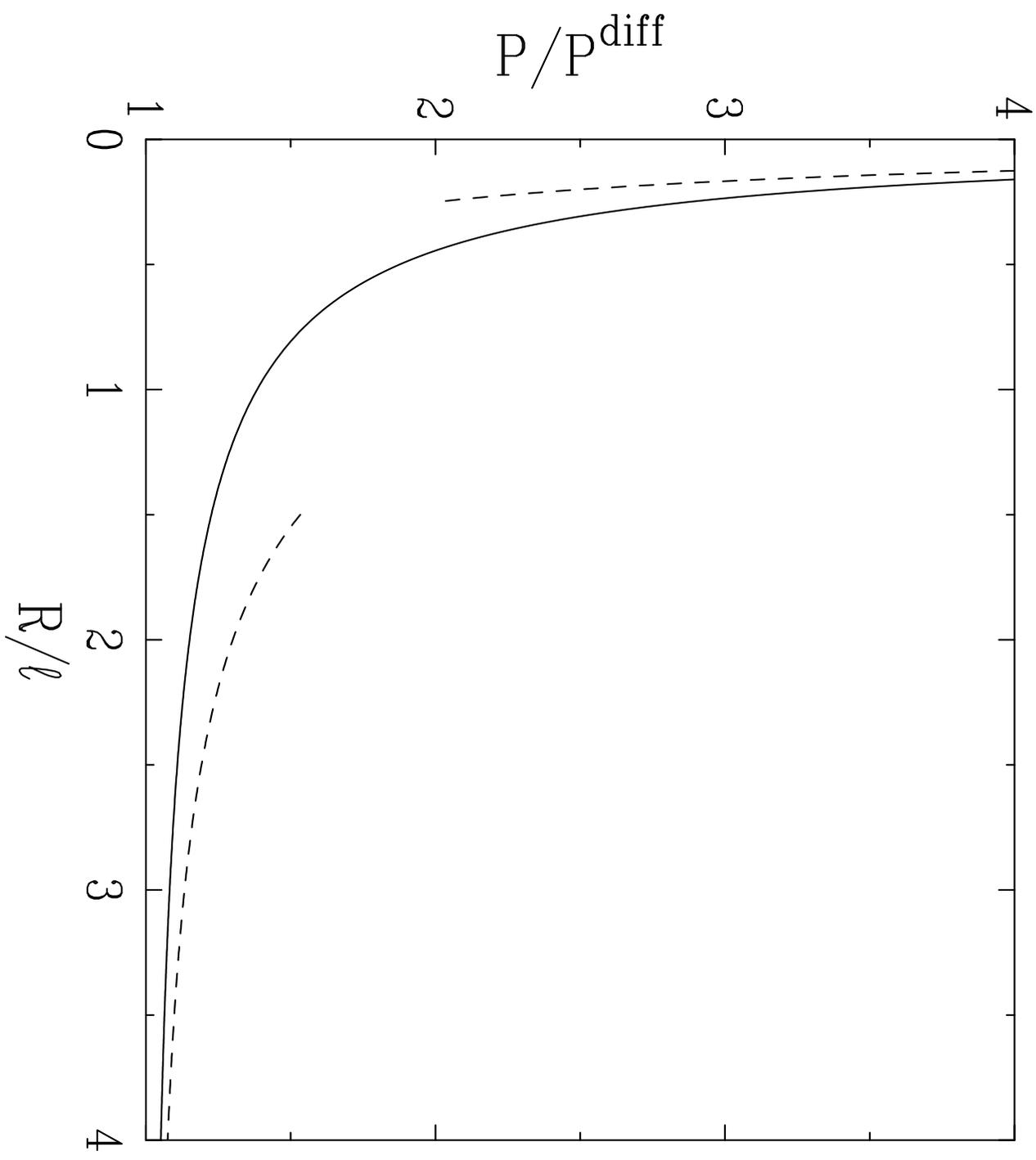
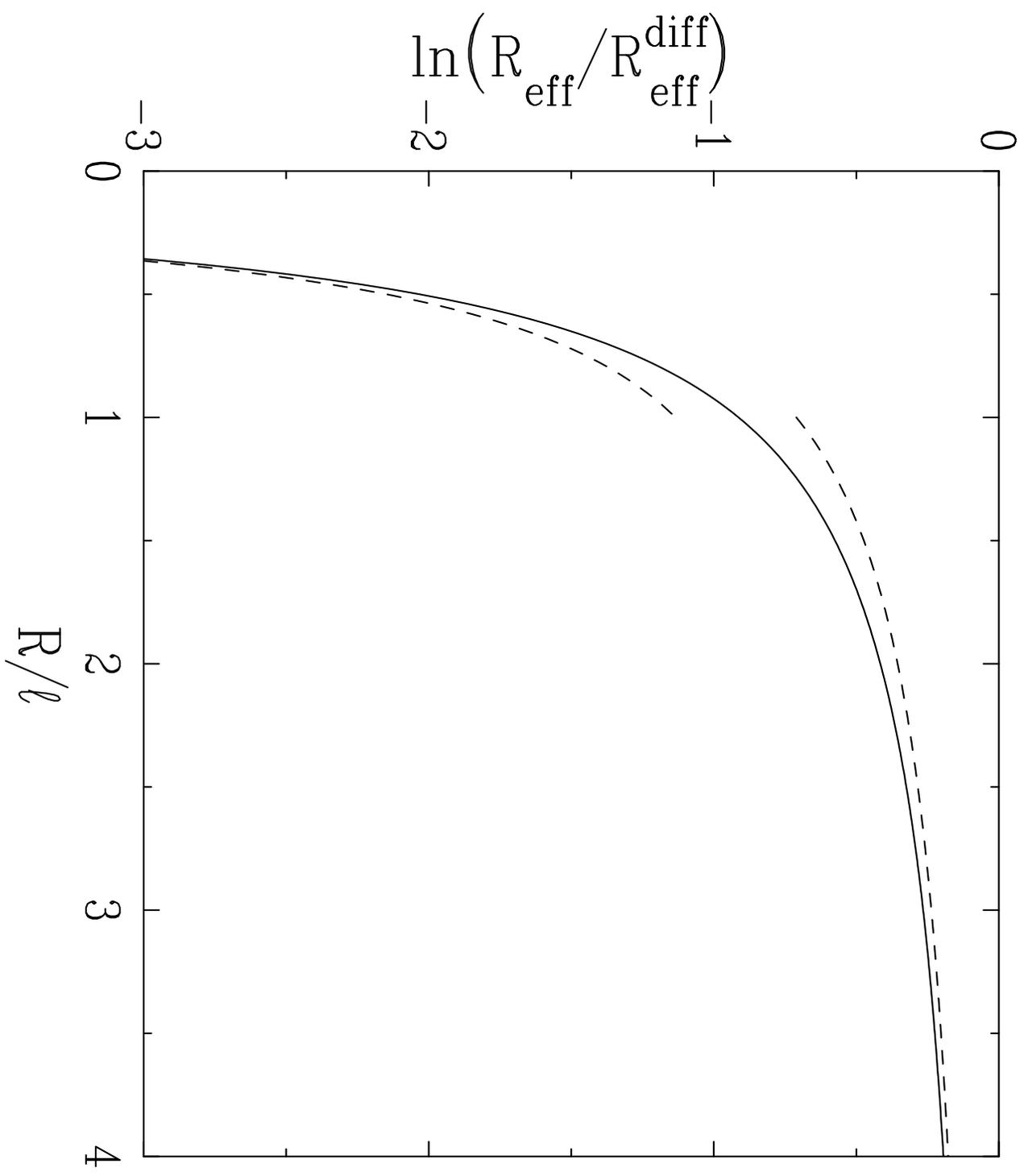
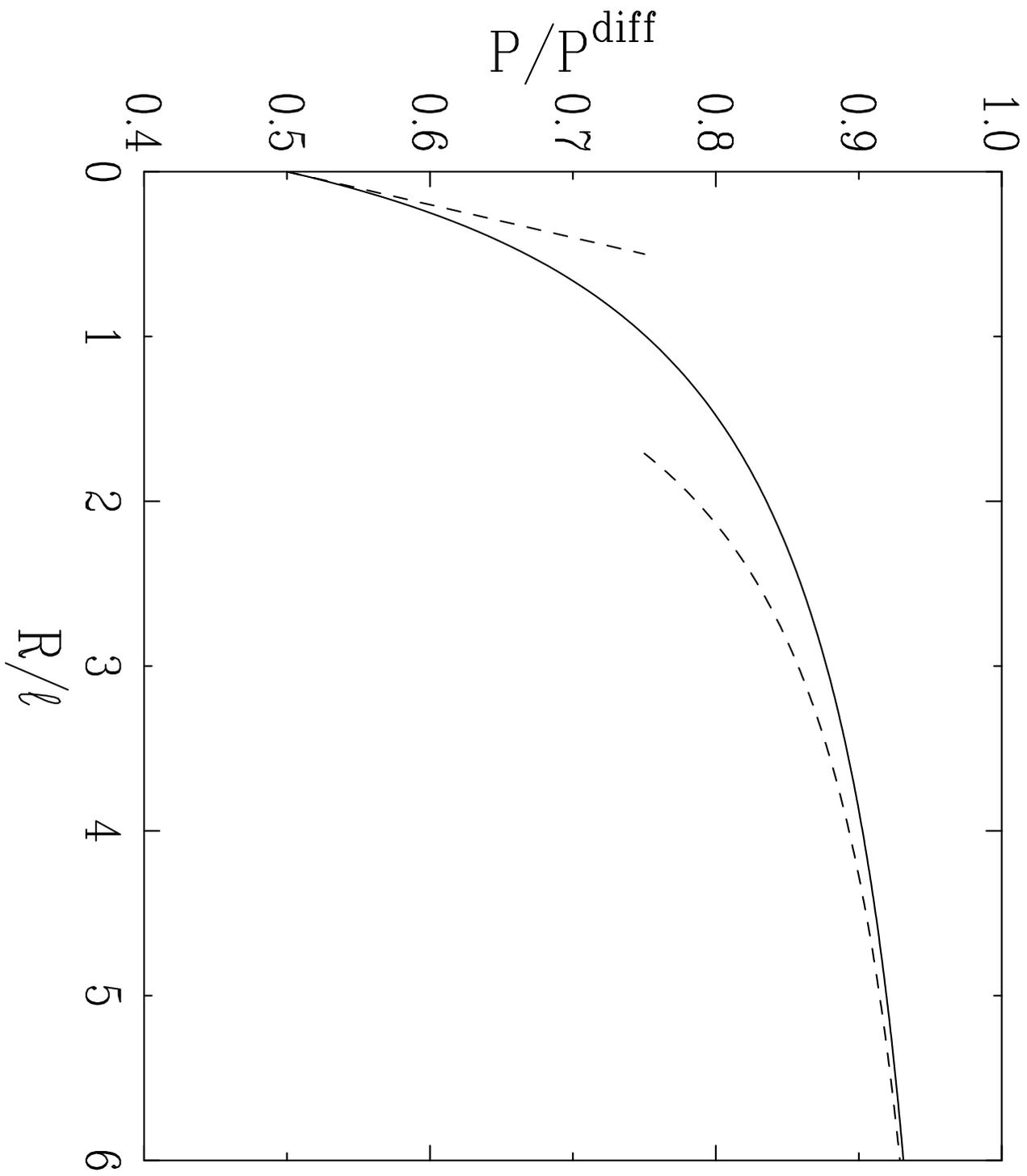
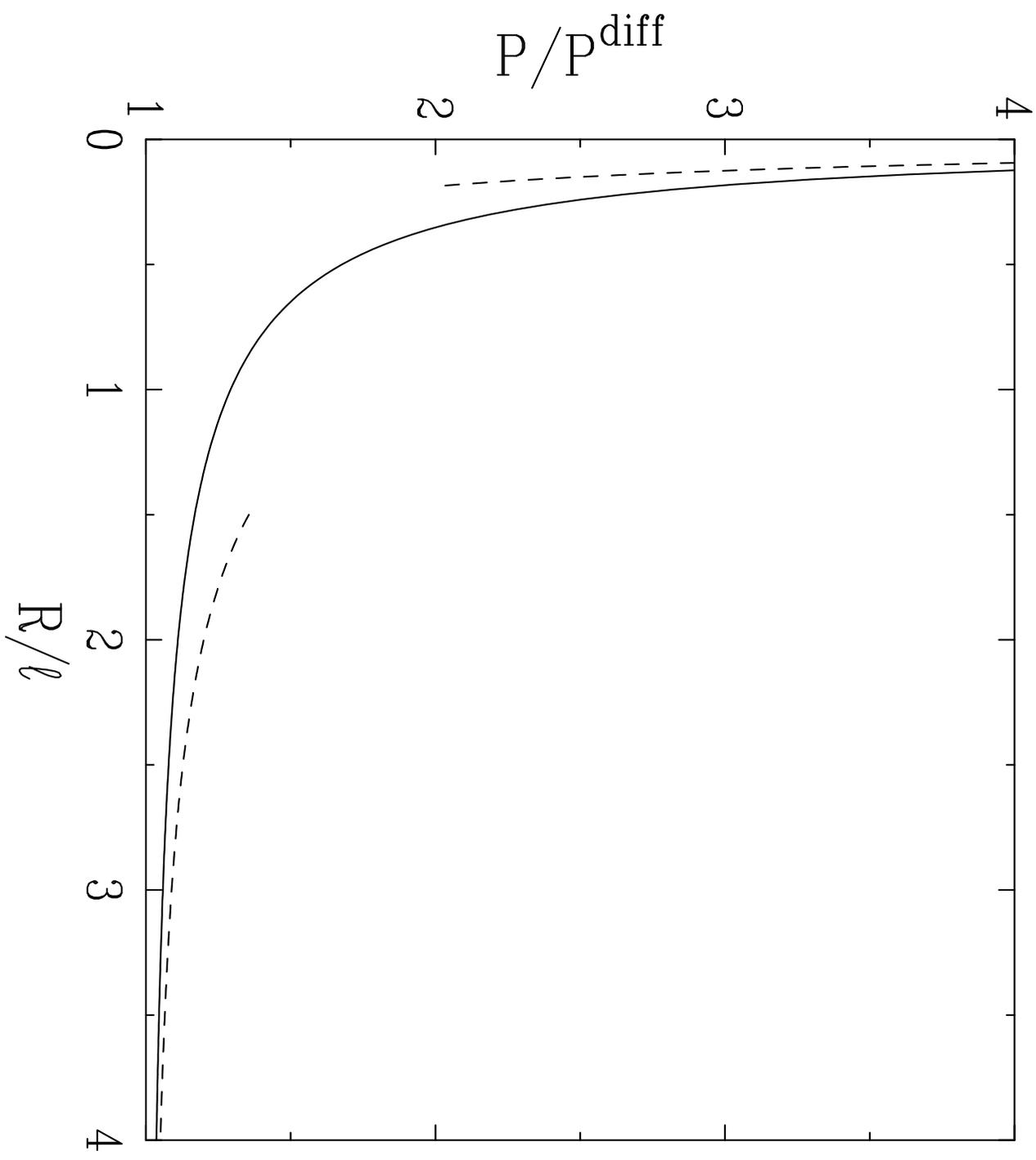
\bye